\documentclass[preprint]{aastex}
\usepackage{spr-astr-addons}
\usepackage{url}
\usepackage{rotating}

\RequirePackage{color}

\begin{document}

\title{Application of the MST clustering to the high energy $\gamma$-ray sky. \\
IV - Blazar candidates found as possible counterparts of photon clusters} 

\shorttitle{MST cluster detection of $\gamma$-ray blazar candidates}
\shortauthors{R. Campana et al.}

\author{R.~Campana\altaffilmark{}}
\affil{INAF/IASF-Bologna, Via Gobetti 101, I-40129, Bologna, Italy.} 
\and 
\author{E.~Massaro}
\affil{INAF/IAPS, via del Fosso del Cavaliere 100, I-00133, Roma, Italy}
\affil{In Unam Sapientiam, Roma, Italy}
\author{E.~Bernieri}
\affil{INFN - Sezione di Roma Tre, via della Vasca Navale 84, I-00146 Roma, Italy.}
\email{campana@iasfbo.inaf.it} 


\begin{abstract}
We present the results of a cluster search in the \emph{Fermi}-LAT Pass 8 
$\gamma$-ray sky by means of the Minimum Spanning Tree algorithm, at energies 
higher than 10 GeV and at Galactic latitudes higher than 25\degr. 
The selected clusters have a minimum number of photons higher than or equal to 5,
a high degree of concentration, and are without a clear corresponding counterpart 
in blazar catalogues. 
A sample of 30 possible $\gamma$-ray sources was obtained.
These objects were verified by applying the standard Maximum Likelihood analysis 
on the \emph{Fermi}-LAT data.
A search for possible radio counterparts in a circle having a radius of 6\arcmin\ 
was performed, finding several interesting objects, the majority of them without 
optical spectroscopical data.
These can be considered as new blazar candidates.
Some of them were already noticed as possible blazars or Active Galactic Nuclei in previous
surveys, but never associated with high energy emission.
These possible counterparts are reported and their properties are discussed. 
\end{abstract}

\keywords{$\gamma$-rays: observations -- $\gamma$-rays: source detection}
  
\section{Introduction }\label{s:introduction}

The \emph{Fermi}-Large Area Telescope (LAT) sky survey at $\gamma$-ray energies has 
shown that blazars constitute the largest class of extragalactic high energy sources 
\citep[see, e.g., the review paper by][]{massaro16}.
For this reason, since a few years the searches for new blazars based on multifrequency 
approaches have been very successful, providing new samples of blazars and candidates 
as, for example,
WIBRaLS \citep{dabrusco14} and 1WHSP \citep{arsioli15}.

\citet[Paper I]{paperI} applied successfully the Minimum 
Spanning Tree (hereafter MST) algorithm for searching new spatial clusters of $\gamma$ 
rays which could be an indication for localized faint high energy sources.
In particular, it was illustrated how MST is 
useful for finding clusters having a small number of photons, but likely related 
to pointlike sources.

In two other previous papers \citep[][hereafter Paper II and III]{paperII,paperIII} 
we reported samples of blazars and blazar candidates associated with photon clusters 
found by means of the MST algorithm.
These sources were included in the 5th Edition of the Roma-BZCAT \citep{massaro14} 
and in the 1WHSP sample.

In this paper we 
extend our MST analysis of the 
\emph{Fermi}-LAT $\gamma$-ray sky at energies higher than 10 GeV,
reporting the discovery of 30 new photon clusters not associated
with known blazars but having interesting radio counterparts with blazar-like 
characteristics at angular distances lower than a few arcminutes.
Some of them where already reported in recent samples of quasar candidates.
Optical spectra are available from the Sloan Digital Sky Survey \citep[SDSS,][]{alam15} 
for four objects and only one from the 6dF survey \citep{jones04,jones09},
therefore for a better classification we considered the mid-infrared photometric 
colours from the WISE \citep[Wide-field Infrared Survey Explorer,][]{wright10}
database according the criteria introduced by \cite{massaro11}
and for the WIBRaLS sample \cite{dabrusco12}.
As in the previous papers, in this search we considered the Fermi-LAT Pass 8 sky
for Galactic latitudes higher than $|25\degr|$.

\section{Photon cluster detection by means of the MST algorithm}

The Minimum Spanning Tree \citep{campana08,campana13} 
is a source-detection algorithm useful for searching clusters in a 
field of points. 
A brief description of MST was presented elsewhere (e.g. in Paper I), and therefore we provide
here only a brief summary of this method.

Consider a two-dimensional set of $N$ \emph{nodes}: one can define a set
$\{\lambda_i\}$ of weighted \emph{edges} connecting them. 
The MST is the unique tree (i.e. a graph without closed loops) that connects all the 
nodes with the minimum total weight, defined as $\min [\Sigma_i \lambda_i]$. 
For a set of points in a Cartesian frame, the edges are the lines joining the nodes and 
the weights are their lengths, while for a region on the celestial sphere the edge 
weights are the angular distances between pairs of photons. 

Once the MST is computed, a set of subtrees corresponding to clusters of
photons is extracted by means of a \emph{primary} selection, consisting of:
\emph{i) separation}: remove all the edges having a length $\lambda > \Lambda_\mathrm{cut}$, the 
separation value, defined in units of the mean edge length $\Lambda_m = (\Sigma_i \lambda_i)/N$, 
obtaining a set of disconnected sub-trees;
\emph{ii) elimination}: remove all the sub-trees having a number of nodes $n \leq N_\mathrm{cut}$, 
leaving only the clusters having a size over a fixed threshold. 
The remaining set of sub-trees provides a first list of candidate cluster and a 
\emph{secondary} selection is applied to extract the most robust candidates for 
$\gamma$-ray sources.
The suitable parameter for this selection \citep{campana13}, also useful for 
evaluating the ``goodness" of the accepted clusters, is the \emph{magnitude} of the cluster:
\begin{equation}
M_k = n_k g_k  
\end{equation}
where $n_k$ is the number of nodes in the cluster $k$ and the \emph{clustering parameter} $g_k$ 
is the ratio between $\Lambda_m$ and $\lambda_{m,k}$, the mean length of the $k$-th cluster edges. 
The probability to obtain a given magnitude value combines that of selecting a cluster with
$n_k$ nodes together with its ``clumpiness'', compared to the mean separation in the field.  
It was found that $\sqrt{M}$ is a good estimator of statistical significance of MST clusters.
In particular, a lower threshold value of $M$ around 15--20 would reject the 
large majority of spurious (low significance) clusters. 

The cluster centroid coordinates are obtained
by means of a weighted mean of the photons' coordinates \citep{campana13}.
If the cluster can be associated with a genuine pointlike 
$\gamma$-ray source, the radius  of the circle centred at the centroid and containing 
the 50\% of photons in the cluster, the {\it median radius} $R_m$,  should be smaller than 
or comparable to the 68\% containment radius of instrumental Point Spread Function 
\citep[PSF, see][]{ackermann13b}.

\begin{table*}
\caption{New MST clusters at $E > 10$ GeV and their possible counterparts. 
Columns 2--7 report the MST parameters of the MST cluster, column 8 gives the 
J2000 positions in the AllWISE catalog and in column 9 the angular distance to the cluster 
centroid is given. 
}
\label{table1}
{
\begin{tabular}{crrccccccl}
\hline
MST cluster & RA~~    & DEC~       & $n$& $g$ &  $M$ & $R_m$ & AllWISE source       & $\Delta \theta$ & Notes  \\
	     & J2000   & J2000     &    &       &       &  deg  &                     &       $'$       &   \\
\hline
MST 0041$-$1608 &  10.433 & $-$16.137 &  5 & 7.040 & 35.198 & 0.042 & J004141.21$-$160747.2 & 0.82 &   \\ 
MST 0059$-$3512 &  14.856 & $-$35.214 &  9 & 3.723 & 33.511 & 0.128 & J005931.47$-$351049.1 & 2.39 &   \\ 
MST 0133$-$4533 &  23.339 & $-$45.558 &  6 & 7.133 & 42.800 & 0.034 & J013309.28$-$453524.0 & 2.80 &   \\ 
MST 0135$+$0257 &  23.790 &     2.951 &  5 & 4.012 & 20.062 & 0.074 & J013507.04$+$025542.6 & 1.53 &   \\ 
MST 0137$+$2247 &  24.491 &    22.793 &  9 & 3.635 & 32.719 & 0.068 & J013801.12$+$224808.7 & 0.96 &   \\ 
MST 0201$-$4348 &  30.272 & $-$43.800 &  5 & 4.052 & 20.260 & 0.073 & J020110.93$-$434655.5 & 1.48 &   \\ 
MST 0207$-$2403 &  31.899 & $-$24.054 &  7 & 3.578 & 25.043 & 0.067 & J020733.39$-$240202.0 & 1.32 &   \\ 
MST 0310$-$1039 &  47.715 & $-$10.657 &  8 & 2.872 & 22.973 & 0.108 & J031034.10$-$103714.9 & 4.83 &   \\ 
MST 0338$-$2850 &  54.719 & $-$28.841 & 14 & 3.389 & 47.443 & 0.089 & J033859.60$-$284619.9 & 4.39 & cnf \\ 
MST 0350$+$0640 &  57.505 &     6.677 &  7 & 4.437 & 31.062 & 0.041 & J034957.83$+$064126.2 & 1.16 & cnf \\ 
MST 0350$-$5146 &  57.605 & $-$51.774 &  6 & 6.662 & 39.970 & 0.059 & J035028.30$-$514454.3 & 1.78 & cnf \\ 
MST 0359$-$0235 &  59.872 &  $-$2.590 &  8 & 3.077 & 24.613 & 0.065 & J035923.48$-$023501.8 & 1.50 &  \\ 
MST 0703$+$6808 & 105.830 &    68.144 &  5 & 4.402 & 22.012 & 0.048 & J070315.90$+$680831.2 & 0.34 & cnf \\ 
MST 0805$+$3835 & 121.488 &    38.589 &  9 & 4.201 & 37.811 & 0.056 & J080551.75$+$383538.0 & 1.07 &  \\ 
MST 1020$+$0510 & 155.011 &     5.167 &  9 & 2.981 & 26.833 & 0.140 & J102015.12$+$050910.6 & 3.23 & cnf rg \\ 
MST 1405$-$1853 & 211.481 & $-$18.892 &  8 & 5.039 & 40.312 & 0.033 & J140545.58$-$185123.8 & 3.15 &  \\ 
MST 1410$+$1438 & 212.623 &    14.643 &  7 & 4.747 & 33.232 & 0.060 & J141028.05$+$143840.2 & 0.36 & rg \\ 
MST 1427$-$1823 & 216.859 & $-$18.399 & 12 & 4.112 & 49.347 & 0.087 & J142725.93$-$182303.7 & 0.88 & cp \\ 
MST 1431$-$3122 & 217.793 & $-$31.372 & 15 & 3.073 & 46.092 & 0.076 & J143109.22$-$312038.8 & 1.71 &  \\
MST 1439$-$2524 & 219.872 & $-$25.415 &  7 & 5.569 & 38.981 & 0.041 & J143934.65$-$252459.1 & 1.22 &  \\ 
MST 1503$+$1651 & 225.852 &    16.862 &  6 & 4.588 & 27.529 & 0.042 & J150316.56$+$165117.6 & 1.93 & cnf \\ 
MST 1514$-$0949 & 228.679 &  $-$9.830 & 12 & 2.764 & 33.172 & 0.102 & J151449.75$-$094838.4 & 2.00 & \\  
MST 1546$-$1002 & 236.569 & $-$10.034 &  9 & 3.191 & 28.720 & 0.078 & J154611.48$-$100326.1 & 1.88 &  \\ 
MST 1547$-$1531 & 236.823 & $-$15.521 &  9 & 4.584 & 41.258 & 0.054 & J154725.68$-$153237.0 & 2.42 & cp SD \\ 
MST 1605$-$1142 & 241.262 & $-$11.706 &  9 & 3.882 & 34.936 & 0.060 & J160517.53$-$113926.8 & 4.62 & \\ 
MST 1640$+$0640 & 250.044 &     6.478 &  6 & 3.586 & 21.518 & 0.025 & J164011.05$+$062826.9 & 0.25 & \\ 
MST 1643$+$3317 & 250.943 &    33.293 & 10 & 3.054 & 30.540 & 0.055 & J164339.46$+$331647.8 & 1.62 & cnf \\ 
MST 1716$+$2307 & 259.034 &    23.129 &  7 & 5.267 & 36.871 & 0.027 & J171603.22$+$230822.9 & 1.31 &  \\ 
MST 2037$-$3836 & 309.436 & $-$38.607 &  5 & 4.192 & 20.958 & 0.061 & J203733.38$-$383635.8 & 2.19 &  \\ 
MST 2245$-$1734 & 341.389 & $-$17.575 &  6 & 3.405 & 20.432 & 0.044 & J224531.85$-$173358.9 & 0.64 &  \\ 
\hline
\end{tabular}
}
~\\
Notes: \\ cnf: possible counterpart confusion; \\
rg: possible radio galaxy; \\
cp: close pair; \\
SD; coordinates from SDSS.
\end{table*}

\section{The MST cluster populations}

As reported in Papers II and III, \emph{Fermi}-LAT data (Pass 8) above 10~GeV, covering the 
whole sky in the 7 years time range from the start of mission (2008 August 04) 
up to 2015 August 04, were downloaded from the FSSC 
archive\footnote{\url{http://fermi.gsfc.nasa.gov/ssc/data/access/}}.
Standard cuts on the zenith angle and data quality were applied.

We searched for cluster of $\gamma$ photons by means of MST in the sky after the
exclusion of the Galactic belt up to a latitude $|b| \leq 25^{\circ}$ to reduce the
possibility of finding clusters originated by local high background fluctuations.
Each of these two broad regions was then divided into ten smaller parts
where MST was applied.
Regions were selected with 2\degr\ width overlapping strips along their boundaries
to avoid missing clusters; in the case of multiple detections the cluster with the highest $M$ 
was inserted in the sample.
The parameters of primary selection of clusters were $N_\mathrm{cut} = 4$
and $\Lambda_\mathrm{cut} = 0.7\,\Lambda_{m}$; then a secondary selection was applied with 
$ M > 20$, that according to the results of \citet{campana13} gives
a very low probability to select spurious clusters. 

A sample of 919 clusters was obtained, of which 716 have a firm 3FGL counterpart.
For 165 1FHL counterparts were found, five of which not in the 3FGL catalogue, 
From the residual sample of 198 clusters we then sorted out 16 clusters that were found
associated with 1WHSP sources (Paper II), 35 other clusters very close to
blazars in the 5BZCAT catalogue (Papers I and III), 5 corresponding to 
$\gamma$-ray sources in the recent 2FHL catalogue at energies higher than 50 GeV 
\citep{2fhl} and 1 corresponding to the GRB~130427A \citep{maselli14,ackermann14}.
Thus, the final sample contains 141 clusters not related to previously known $\gamma$-ray
sources or blazars.
 
From the latter sample, a further subselection was made by applying the following criteria: 
$i)$ $M > 25$, or $ii)$ $g > 3$.
Note that the latter criterium is particularly strong, 
because it selects clusters with a high photon concentration, as expected from point-like
sources, while extended features or close cluster pairs are more efficiently rejected, as already
discussed in \cite{paperIII}. 
It should be noticed that, however, sources with $g \le 3$ have been shown to be 
sometimes significant.
Then, the ASDC sky explorer tool\footnote{\url{http://www.asdc.asi.it/}} was used
to search within a region of a radius of 6\arcmin, that in \cite{paperI} has 
been shown to be optimal for cross-matching, for radio sources from a large set of catalogues, 
including 
NVSS \citep{condon98},
FIRST \citep{white97},
SUMSS 2.1 \citep{mauch03},
PMN  \citep{gregory94},
and many others. 
We considered only sources with possible radio and optical counterparts and obtained a final
list of 30 objects which 
could be considered for a deeper analysis to assess their likely status as blazar candidates.
The optical counterparts of the radio sources have been searched using several databases
accessible from the ASDC explorer tool, connected to $Vizier$ and thus including SDSS, USNO and 
COSMOS catalogues,
and considering the uncertainty on the radio source location.
All these sources had also a clear infrared counterpart in the AllWISE \citep{cutri13} catalogue 
and the positional correspondence was verified in the 
images\footnote{\url{http://irsa.ipac.caltech.edu/applications/wise/}}. 
Table~\ref{table1} reports the main MST parameters of these clusters and the coordinates
of the corresponding possible counterparts in the AllWISE catalogue after the accurate
verification in multiband IR images.
When within the region of interest of some clusters 
more than one radio source was found, and when the selection of the best radio counterpart
(also using literature data) was not safe, the note `cnf' (possible confusion) is added.

Several sources were already 
reported as blazar candidates in WIBRaLS \citep[][marked as WBR in the following
Table~\ref{table2}]{dabrusco14} 
or as QSO by \citet[][B15 in Table~\ref{table2}]{brescia15}, 
or their flat radio spectrum was confirmed by low frequency data 
\citet[][M14 in Table~\ref{table2}]{massaro14};
one is in the 1WHSP sample (1WH in Table~\ref{table2}), but not reported in Paper II because 
the cluster has 5 photons, lower than the minimum number considered in that paper.
Only for one source we found a possible correspondence with a high energy source
reported in the 3EG catalogue \citep{3eg}, but up to now undetected in the Fermi sky.

\section{$TS$ evaluation from ML analysis}

We performed also a standard unbinned likelihood analysis for each MST cluster. 
A Region of Interest (ROI) of 10\degr\ radius was selected around the MST 
cluster centroid, and standard screening criteria were applied to the 
\emph{Fermi}-LAT data above 3 GeV. 
The likelihood analysis was performed considering all the 3FGL sources within 
20\degr\ from the cluster centroid, as well as the Galactic and extragalactic 
diffuse emission. 
A further source with a power-law spectral distribution was assumed at the MST 
coordinates. 
The normalization and spectral index of all the 3FGL sources within the ROI 
was allowed to vary in the fitting, while the parameters of the sources between 
10\degr\ and 20\degr\ from the center of the field of view were fixed to their 
catalogue values.
From this analysis, we derived the likelihood Test Statistics ($TS$) and fluxes 
in the two 3--300 GeV and 10--300 GeV bands. 
These results are reported in Table~\ref{table2} together with some other photometric
and spectroscopic data, when available.

We found only five sources with $\sqrt{TS} < 5$, but one (MST 0041$-$1608) is
almost borderline ($\sqrt{TS} = 4.96$) and could be considered confirmed.
Two of the remaining four clusters, MST 0310$-$1039 and MST 2245$-$1734, have 
$\sqrt{TS} > 4.5$, that correspond to a detection significance above 
roughly 4.5\,$\sigma$, therefore can be reasonably considered as safe (although 
lower than the conventional 5\,$\sigma$ threshold to be included in the FGL source
catalogs), whereas the other 
two (MST 0201$-$4348 and MST 2037$-$3836), with ML significances of 3.5 and 2.3, 
respectively, are questionable and their actual detection will be discussed 
in detail the following section.

\section{Properties of new blazar candidates}

In Table~\ref{table2} we reported in addition to ML significance and spectral data
some other parameters of interest: the radio flux density at 1.4 GHz
from NVSS or FIRST, optical magnitude in the $F$ (photographic red band magnitude 
$R$ from The Guide Star Catalog, Version 2.3.2 (GSC2.3, STScI, 2006), or $r$ 
(from SDSS) bands and spectroscopic information or the redshift 
$z$, useful for the understanding of the main properties of these blazar candidates.

\begin{figure}[h]
\centering 
\includegraphics[width=0.48\textwidth]{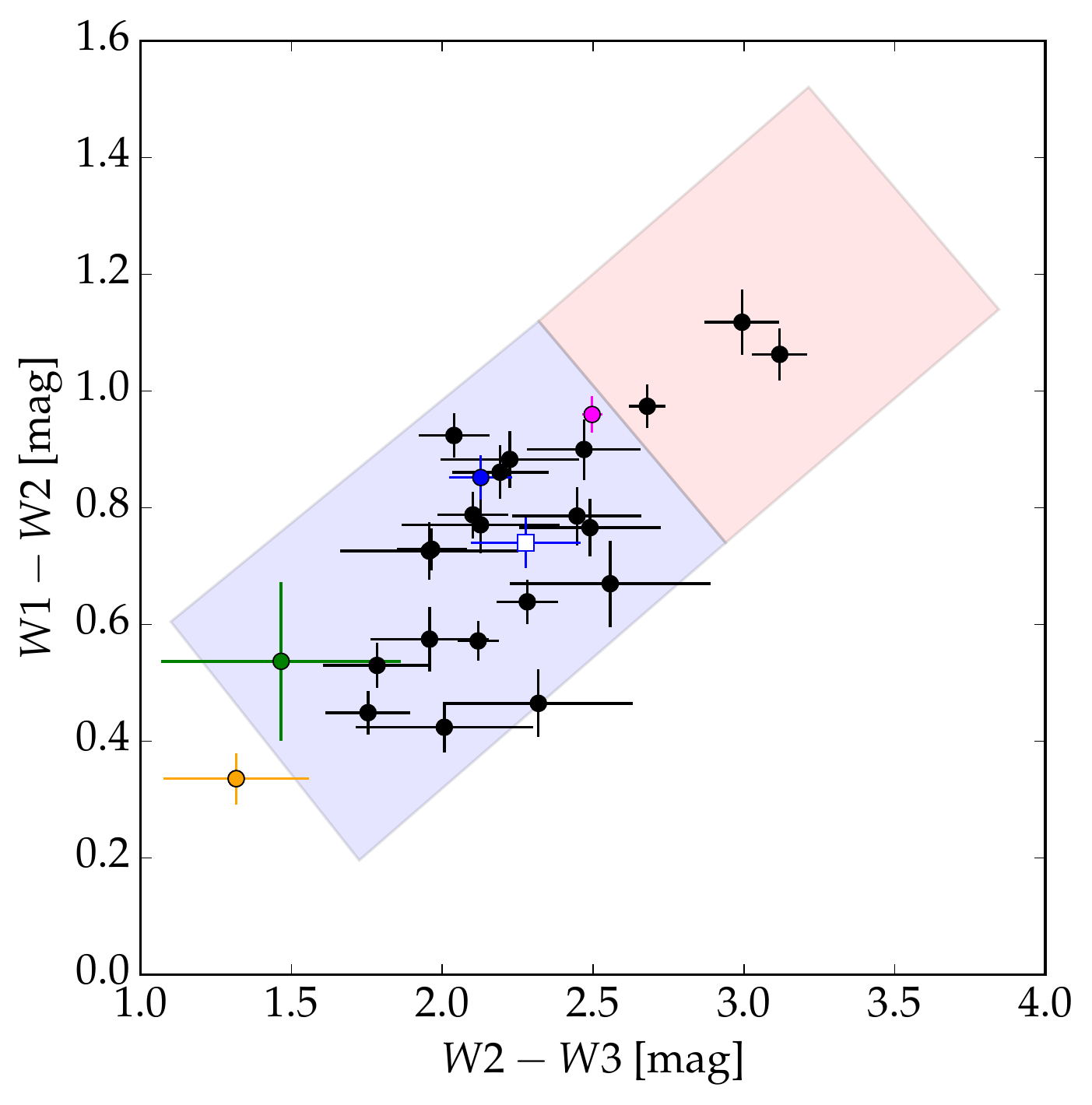}
\caption{Plot of the infrared colours of the MST-selected BL Lac objects from the AllWISE catalogue.
The shaded regions define the WISE Blazar locus where are present $\gamma$-ray loud blazars. 
The blue-shaded region represents the locus where there is a concentration of BL Lac objects, 
while the red-shaded region correspond to FSRQ objects.
The green point corresponds to the second counterpart of MST 0350$+$0640; the magenta point 
to the counterpart of MST 1431$-$3122 and the orange point, just outside the blazar region,
is that of MST 1439$-$2524.
The blue circle and the open blue square indicate the colours of the WISE candidate couterparts to 
MST 1546$-$1002 and MST 1547$-$1531, respectively, as explained in the text.
}
\label{f1}
\end{figure}

As apparent in Table~\ref{table1} all sources have likely AllWISE counterparts and
for 25 of them photometric data are given in the $W1$ (13.4~$\mu$m), 
$W2$ (4.6~$\mu$m) and $W3$ (12~$\mu$m) bandpasses (Table~\ref{table3}).
Further 13 of them are also detected in the $W4$ (22~$\mu$m) filter.
\citet{massaro13} investigated the nature of unidentified 2FGL sources
in the two WISE colour plot by means of the Kernel Density Estimator
replacing the constraint on the detection at 22~$\mu$m with the occurrence
of an associated radio emission.
Considering that our candidates are all detected in the radio band, we 
computed their two WISE colour plot that is shown in Figure~\ref{f1}.
No reddening correction was applied because at the Galactic latitudes of
our sources its largest effect on the colours was of only a few hundredths of
magnitude, lower than typical uncertainties.  
In this plane $\gamma$-ray blazars are essentially concentrated within the 
two coloured areas,
defined in the figures reported in \citet{massaro13}, \citet{dabrusco14},
and \citet{massarodabrusco16}:
BL Lac objects are mainly concentrated in the blue area, while FSRQ are mostly 
found in the red one.
In their data there is no definite boundary between BL Lac and FSRQ regions, and 
the given separation is only indicative.
All our candidates have colours matching very well this \emph{locus}, 
with a clear dominance for the BL Lac region; there is only one source (MST 1439$-$2524)
just outside the region, but its distance from the boundary is within 1$\sigma$, 
not large enough to be significant.
Note, in particular, the very close similarity between our plot and that given in
Figure 2 of \citet{massaro13}.

For what concerns the 13 sources with also $W4$ data, 5 of them, typically those with
the highest signal to noise ratio, were also reported in WIBRaLS (Table~\ref{table2}) and therefore are
located in the blazar locus in the 3 colour plot \citep{dabrusco14}. 
Three other sources lie also in the same region, whereas all the remaining 5 have uncertainties on $W4$ of $\approx$0.4 mag and are
compatible with the expected position within one standard deviation.

In the following we discuss the individual properties of the candidates.

\begin{table*}[htb!]
\caption{Standard unbinned likelihood analysis of the \emph{Fermi}-LAT data, see 
Sect. 4 for details. 
The columns 3--4 report photon fluxes in units of 
10$^{-11}$ ph\,cm$^{-2}$\,s$^{-1}$.  
For sources below the usual significance threshold ($\sqrt{TS}=5$) only upper 
limits are given.
Radio flux densities at 1.4 or 0.8 GHz, optical magnitude, RASS X-ray fluxes (0.1--2.4 keV) 
in erg cm$^{-2}$~s$^{-1}$ and the redshift, when available, are given in the colums from 6 to 9. 
}
\label{table2}
\centering
\scriptsize{
\begin{tabular}{lccccrcccl}
\hline
 MST cluster &  $\sqrt{TS}$  &    Flux      &    Flux     & Photon   &  $F_{1.4/0.8}$ &  $F/r$  &  $F_X$~  & z & Notes \\
             &               &  3--300 GeV  & 10--300 GeV &      index         &    mJy     &  mag  & $\times10^{-12}$     &   \\
\hline
MST 0041$-$1608 & 5.0 & $3.4\pm1.5 $ & $1.3\pm0.7$ & $1.7\pm0.4$ & 18 & 18.8 &  ---  & ---  &  \\ 
MST 0059$-$3512 & 7.0 & $7.8\pm2.2 $ & $2.0\pm0.9$ & $2.1\pm0.3$ & 79 & 18.0 & 1.93  & ---  &  6dF \\ 
MST 0133$-$4533 & 5.4 & $4.7\pm1.7 $ & $1.4\pm0.7$ & $2.0\pm0.4$ & 16 & 19.5 & 2.56  & ---  &  s   \\ 
MST 0135$+$0257 & 5.3 & $4.5\pm1.8 $ & $2.4\pm1.2$ & $1.4\pm0.9$ & 22 & 19.0 & 0.38  & ---  & r gal B15 \\ 
MST 0137$+$2247 & 7.9 & $11.0\pm2.7$ & $3.3\pm1.2$ & $2.0\pm0.3$ & 42 & 18.7 & ---   & ---  & r WBR B15 \\ 
MST 0201$-$4348~* & 3.5 & $< 3.1$    & $< 0.3$     &      ---    & 45 & 19.8 & 1.53  & ---  & s XM 1WH \\ 
MST 0207$-$2403 & 6.4 & $7.3\pm2.2$ & $2.3\pm1.0$  & $1.9\pm0.3$ & 186& 19.5 & ---   & ---  & WBR \\ 
MST 0310$-$1039 & 4.5 & $< 6.2$     & $< 1.3$      &     ---     & 80 & 18.1 & ---   & ---  &  \\
MST 0338$-$2850 & 8.0 & $7.0\pm2.4$ & $2.8\pm1.0$ & $1.7\pm0.2$  & 42 & 17.3 & ---   & ---  &  \\ 
MST 0350$+$0640 & 6.7 & $5.6\pm1.8$ & $3.1\pm1.2$ & $1.4\pm0.7$  &  9 & ---  & ---   & ---  & ($a$) \\ 
~~              & --- & ---         &    ---      &    ---       & 17 & 17.4 & 1.48  & ---  & ($b$)  \\ 
~~              & --- & ---         &    ---      &    ---       & 158& 19.9 & ---   & ---  & ($c$)  \\ 
MST 0350$-$5146 & 5.8 & $5.6\pm2.0$ & $2.3\pm1.0$ & $1.7\pm0.3$  & 11 & 17.5 & 9.11  & ---  & s \\ 
MST 0359$-$0235 & 5.8 & $7.3\pm2.3$ & $3.0\pm1.2$ & $1.7\pm0.3$  & 14 & 18.6 & ---   & ---  &  \\ 
MST 0703$+$6808 & 5.1 & $3.8\pm2.4$ & $1.9\pm1.1$ & $1.5\pm1.0$  & 30 & 17.6 & ---   & ---  & M14 \\ 
MST 0805$+$3835 & 7.1 & $7.9\pm2.1$ & $1.8\pm0.8$ & $2.2\pm0.3$  & 13 & 20.4 & ---   &  nl  & r SD WBR \\ 
MST 1020$+$0510 & 6.3 & $9.0\pm2.4$ & $2.0\pm1.0$ & $2.2\pm0.4$  &  9 & 19.9 & ---   &  nl  & r ($d$) \\ 
~~              & --- &    ---      &    ---      &    ---       &  2 & 21.4 & ---   & ---  & r ($e$) \\ 
~~          	  & --- &    ---      &    ---      &    ---       & 10 & 18.1 & ---   & 0.276 & r gal SD ($f$) \\ 
MST 1405$-$1853 & 5.7 & $5.4\pm2.0$  & $2.1\pm1.2$ & $1.8\pm0.3$ & 29 & 18.6 & ---   &  --- &  \\ 
MST 1410$+$1438 & 8.1 & $10.\pm2.4$  & $2.7\pm1.0$ & $2.1\pm0.3$ &434 & 16.7 & ---   & 0.144 & r SD el WBR \\ 
MST 1427$-$1823 & 7.4 & $11.\pm2.7$  & $2.8\pm1.0$ & $2.1\pm0.3$ & 48 & 18.1 & ---   &  --- &  \\ 
MST 1431$-$3122 & 7.9 & $12.\pm2.8$  & $3.1\pm1.2$ & $2.1\pm0.3$ &254 & 18.1 & 0.27  & --- & CRT WBR \\
MST 1439$-$2524 & 6.7 & $8.3\pm2.4$  & $2.6\pm1.1$ & $1.9\pm0.4$ & 35 & 16.3 & ---   &  --- &  \\ 
MST 1503$+$1651 & 5.5 & $4.2\pm1.9$  & $1.7\pm0.8$ & $1.7\pm0.4$ & 13 & 18.8 & ---   &   nl & r SD \\ 
MST 1514$-$0949 & 10.4& $26.\pm4.3$  & $3.3\pm1.2$ & $2.7\pm0.3$ &349 & 19.1 & ---   &  --- & CRT  \\  
MST 1546$-$1002 & 6.4 & $10.\pm2.4$  & $2.7\pm1.1$ & $2.1\pm0.3$ & 52 & ---  & ---   &  --- & CRT M14 \\ 
MST 1547$-$1531 & 5.4 & $7.3\pm2.5$  & $2.2\pm1.0$ & $2.0\pm0.4$ & 10 & ---  & ---   &  --- &  \\
MST 1605$-$1142 & 5.9 & $8.0\pm2.6$  & $3.0\pm1.1$ & $1.8\pm0.3$ &258 & 17.0 & 0.28  &  --- & 3EG CGR WBR M14\\ 
MST 1640$+$0640 & 5.5 & $5.4\pm2.2$  & $1.7\pm0.8$ & $1.9\pm0.4$ & 40 & 18.3 & ---   &  --- &  \\ 
MST 1643$+$3317 & 6.3 & $5.0\pm1.5$  & $2.2\pm0.8$ & $1.6\pm0.1$ & 66 & 19.9 & 0.48  &  --- & r gal B15 \\ 
MST 1716$+$2307 & 9.7 & $13.\pm2.6$  & $3.0\pm1.0$ & $2.2\pm0.2$ &  6 & 19.0 & ---   &  --- & B15 \\ 
MST 2037$-$3836~* & 2.3 & $< 2.7$    & $< 0.3$     &   ---       & 25 & 19.1 & ---   & ---  &  \\  
MST 2245$-$1734 & 4.7 & $< 5.6$      & $< 0.4$     &   ---       & 41 & 19.3 & ---   & ---  & B15 \\ 
\hline
\end{tabular}
}
~\\
Notes:\\ 
* : poorly significant cluster; 
($a$), ($b$), ($c$):  see Section \ref{src10};
($d$), ($e$), ($f$):  see Section \ref{src15};
6dF: spectrum in 6dF survey; 
SD: spectrum in SDSS;
s: radio flux density from SUMSS at 0.8 MHz; 
r: r magnitude from SDSS;
gal: classified as galaxy in SDSS;
nl: no clear emission lines in the spectrum;
B15: QSO candidate for \citet{brescia15}; 
CRT: CRATES flat spectrum radio source; 
CGR: CGRaBS source; 
3EG: possible correspondence with a 3EG source;
WBR: blazar candidate in WIBRaLS \citep{dabrusco14}; 
M14: LORCAT radio source \citet{massaro14};
1WH: blazar candidate in the 1WHSP sample;
XM: X-ray flux from XMMM database.
\end{table*}

\begin{table*}[htb!]
\caption{Infrared colours of the likely counterparts from the AllWISE photometric database.
See Sections~\ref{src10} and \ref{src15} for discussion about sources $(b)$, $(d)$, $(e)$ and $(f)$.}
\label{table3}
\centering
{
\begin{tabular}{ccccl}
\hline
MST cluster & AllWISE counterpart &  $W1 - W2$   &   $W2 - W3$   &  Notes \\
           &          &              &               &         \\
\hline
MST 0041$-$1608 & J004141.21$-$160747.2 &  $0.416 \pm 0.055$  &   --                & \\
MST 0059$-$3512 & J005931.47$-$351049.1 &  $0.424 \pm 0.043$  &  $2.007 \pm 0.295$  & \\
MST 0133$-$4533 & J013309.28$-$453524.0 &  $0.766 \pm 0.049$  &  $2.490 \pm 0.236$  & \\
MST 0135$+$0257 & J013507.04$+$025542.6 &  $0.364 \pm 0.060$  &   --                & \\
MST 0137$+$2247 & J013801.12$+$224808.7 &  $0.729 \pm 0.036$  &  $1.965 \pm 0.116$  & \\
MST 0201$-$4348 & J020110.93$-$434655.5 &  $0.465 \pm 0.058$  &  $2.319 \pm 0.313$  & \\
MST 0207$-$2403 & J020733.39$-$240202.0 &  $0.974 \pm 0.037$  &  $2.680 \pm 0.061$  & \\
MST 0310$-$1039 & J031034.10$-$103714.9 &  $0.575 \pm 0.044$  &  $1.958 \pm 0.197$  & \\
MST 0338$-$2850 & J033859.60$-$284619.9 &  $0.449 \pm 0.037$  &  $1.754 \pm 0.141$  & \\   
MST 0350$+$0640 & J034957.83$+$064126.2 &  $0.537 \pm 0.136$  &  $1.466 \pm 0.397$  & ($b$) \\     
MST 0350$-$5146 & J035028.30$-$514454.3 &  $0.639 \pm 0.038$  &  $2.282 \pm 0.103$  & \\
MST 0359$-$0235 & J035923.48$-$023501.8 &  $0.670 \pm 0.074$  &  $2.557 \pm 0.333$  & \\
MST 0703$+$6808 & J070315.90$+$680831.2 &  $0.530 \pm 0.039$  &  $1.784 \pm 0.179$  & \\
MST 0805$+$3835 & J080551.75$+$383538.0 &  $0.924 \pm 0.038$  &  $2.039 \pm 0.117$  & \\
MST 1020$+$0510 & J101948.28$+$051328.9 &  $0.883 \pm 0.049$  &  $2.224 \pm 0.230$  & ($d$) \\
MST 1020$+$0510 & J102011.75$+$050635.5 &  $0.344 \pm 0.111$  &   --                & ($e$) \\
MST 1020$+$0510 & J102015.12$+$050910.6 &  $0.234 \pm 0.073$  &   --                & ($f$) \\
MST 1405$-$1853 & J140545.58$-$185123.8 &  $0.861 \pm 0.046$  &  $2.192 \pm 0.160$  & \\
MST 1410$+$1438 & J141028.05$+$143840.2 &  $0.572 \pm 0.034$  &  $2.119 \pm 0.068$  & \\
MST 1427$-$1823 & J142725.93$-$182303.7 &  $0.788 \pm 0.040$  &  $2.102 \pm 0.118$  & \\
MST 1431$-$3122 & J143109.22$-$312038.8 &  $0.960 \pm 0.031$  &  $2.497 \pm 0.034$  & \\
MST 1439$-$2524 & J143934.65$-$252459.1 &  $0.336 \pm 0.044$  &  $1.317 \pm 0.241$  & \\
MST 1503$+$1651 & J150316.56$+$165117.6 &  $0.786 \pm 0.050$  &  $2.447 \pm 0.214$  & \\
MST 1514$-$0949 & J151449.75$-$094838.4 &  $1.118 \pm 0.056$  &  $2.994 \pm 0.124$  & \\
MST 1546$-$1002 & J154611.48$-$100326.1 &  $0.852 \pm 0.038$  &  $2.128 \pm 0.104$  & \\
MST 1547$-$1531 & J154725.68$-$153237.0 &  $0.740 \pm 0.043$  &  $2.277 \pm 0.183$  & \\
MST 1605$-$1142 & J160517.53$-$113926.8 &  $1.063 \pm 0.045$  &  $3.119 \pm 0.092$  & \\
MST 1640$+$0640 & J164011.05$+$062826.9 &  $0.726 \pm 0.049$  &  $1.957 \pm 0.295$  & \\
MST 1643$+$3317 & J164339.46$+$331647.8 &  $0.444 \pm 0.048$  &   --                & \\ 
MST 1716$+$2307 & J171603.22$+$230822.9 &  $0.900 \pm 0.052$  &  $2.470 \pm 0.188$  & \\
MST 2037$-$3836 & J203733.38$-$383635.8 &  $0.148 \pm 0.056$  &   --                & \\ 
MST 2245$-$1734 & J224531.85$-$173358.9 &  $0.771 \pm 0.049$  &  $2.127 \pm 0.262$  & \\
\hline
\end{tabular}
}
\end{table*}

\subsection{Comments to individual selected counterparts to MST clusters}

\subsubsection{MST 0041$-$1608}\label{src1}  

This cluster with a bordeline ML significance has only a possible radio counterpart corresponding
to a rather faint object for which few IR and optical photometric data are available. 
The colours are not in conflict with a possible blazar nature of this object.

\subsubsection{MST 0059$-$3512}\label{src2}  

In the searching circle of this cluster there are four radio sources: the selected one is the
brightest, whereas the other fainter radio sources do not have optical counterparts.
It is reported in the RASS-6dFGS catalogue of X-ray selected AGN from the 6dF Galaxy Survey
by \citet{mahony10} with a featureless spectrum characterized by a large blue excess.
There is therefore evidence for considering it a HBL 
\citep[High energy peaked BL Lac,][]{padovani95} object.

\subsubsection{MST 0133$-$4533}\label{src2}  

Two very close optical galaxy-like objects of similar brightness correspond to the SUMSS radio 
source that has a bright counterpart in the RASS \citep{voges99}.
In the WISE two-colour plot it is close to the central line of the strip.
This source is also reported by \citet{mahony10} and in 6dF, but its spectrum is not available.
There is a strong possibility that one of the two objects could be a BL Lac, but a further 
spectroscopic investigation will be useful to confirm its nature. 

\subsubsection{MST 0135$+$0257}\label{src4}  

This object of galaxy type in SDSS was reported as the counterpart of the X rays source 
1RXS J013506.7$+$025\-558 by \citet{mickaelian06} and was indicated as QSO candidate 
by \citet{brescia15}.
No optical spectral data are available.

\subsubsection{MST 0137$+$2247}\label{src5}  

The proposed counterpart is a flat spectrum radio source and blue starlike object in SDSS, 
classified as C class  BL Lac object candidate
by \citet{dabrusco14} and candidate QSO by \citet{brescia15}.

\subsubsection{MST 0201$-$4348}\label{src6}  

This source is listed as blazar candidate in the 1WHSP catalogue and it is detected 
in the radio, IR and X-ray bands.
The lack of optical spectroscopy does not allow a clear classification.

ML analysis gives a significance for this cluster lower than the acceptance threshold and 
therefore we extended the MST analysis to the low energy ranges.
It resulted that the cluster is found only above 10 GeV: a search at energies
higher than 7 GeV gave a cluster at the same position, again with 5 photons but with
a lower $M$ value because of the shorter mean distance between photons; the same
result was obtained at energies higher than 3 GeV, and no cluster resulted in the range
3--7 GeV.
We can conclude that, considering its $M$ value just above the threshold, it is possible 
that this cluster could be originated by a density fluctuation of the high energy background. 
However, the possibility that its emission can be detected only in this range 
cannot in principle be excluded.

\subsubsection{MST 0207$-$2403}\label{src7}  

This flat spectrum radio source is in the CRATES and in the WIBRaLS catalogues.
An optical spectrum by \citet{titov13} has a too low S/N ratio for a safe classification.  

\subsubsection{MST 0310$-$1039}\label{src8}  

Marginal flat spectrum radio source ($\alpha_r \approx -0.4$) with an optical counterpart
without spectral data.
The WISE colours are close to the central line of the blazar region in Figure~\ref{f1}.

\subsubsection{MST 0338$-$2850}\label{src9}  

There are four radio sources in the interesting field: two without optical counterparts, and 
another corresponding to a very faint object.
More interesting appears the brightest radio source with a safe relatively bright optical 
counterpart and likely associated with the 2RXP J033900.0-284621 source (Flesch 2010)
for which no X-ray flux was found in the data archives. 
It presents an extended galaxy image and a very close fainter source not well resolved
in the WISE images, but the colours place it close to the midline of the blazar area.
No spectral data are available for a good classification.

\subsubsection{MST 0350$+$0640}\label{src10}  

There are three radio sources within a 2\arcmin\ distance to cluster centroid and therefore
the possibility of counterpart confusion cannot be excluded.
All these three sources are reported in Table~\ref{table2}.
The radio source NVSS J035006+064209 (note $a$ in Table~\ref{table2}) does not have an optical 
counterpart in POSS; there is a very faint source in AllWISE at an angular distance 
of 9\farcs6 from the radio position.
The other source, NVSS J034957+064126 (note $b$ in Table~\ref{table2}) has a clear optical-IR 
counterpart and was reported as a QSO or a BL Lac object because of the nearby presence 
of the RASS source RX J0350.0+0640, reported as a photometrically variable with a featureless 
continuum source \citep{appenzeller98} and therefore classified as a BL Lac object.
It is also close to the galaxy cluster A0465.
The XRT image confirms the presence of the very close source 1SXPS J034957.6+064126 \citep[1st 
Swift-XRT Point Source catalogue,][]{evans14}.
Its WISE colours place it in the blazar region (see green point in Figure~\ref{f1}).
The third and brightest source, NVSS J035006+064107 (PMN J0350+0641, ATZ A073 
in NED, note $c$ in Table~\ref{table2}) has a steep radio spectrum ($\alpha_r \approx -0.7$).
No optical counterpart is detectable in POSS. 
NED reports the same note by \citet{appenzeller98} despite the radio to X-ray
angular distance is about 1\farcm4 and no X-ray source is detected at its coordinates
in the XRT image.
The second source appears therefore to be the most likely counterpart.

\subsubsection{MST 0350$-$5146}\label{src11} 

There are three SUMSS sources in the search region, of which only the closest one 
to the cluster centroid has optical-IR counterpart.
This object was also reported by \citet{mahony10} in their RASS-6dFGS catalogue
of X-ray selected AGN, but no spectroscopic data are available.
It has a remarkably flat X-ray spectrum with an energy spectral index close to $\approx -0.25$.
This object was already reported as a possible BL Lac object in the early ROSAT associations
\citep[e.g.][]{schwope00}
but never confirmed by optical spectroscopy.
The detection of a $\gamma$-ray counterpart suggests it as an HBL object.

\subsubsection{MST 0359$-$0235}\label{src12} 

The only available data on this radio source are from optical and IR photometric
catalogues and its possible blazar nature is indicated by WISE colours;
optical spectroscopic data are necessary to support this identification.

\subsubsection{MST 0703$+$6808}\label{src13} 

There are only photometric data from optical and IR surveys, the former ones suggesting a 
possible galaxy dominance, but in the WISE colour plot it is close to the midline of the
BL Lac region.
The radio source is reported in the LORCAT catalogue \citep{massaro14} because its
low-frequency radio spectral index is 0.21.
There are two other fainter NVSS radio sources within a radius of $\sim$ 4\arcmin\ that could be
associated with galaxies.
More data are necessary for a safe classification.

\subsubsection{MST 0805$+$3835}\label{src14} 

This source was already indicated as QSO by \citet{dabrusco09} and as a C class 
$\gamma$-ray candidate in WIBRaLS \citep{dabrusco14}
SDSS photometry $r = 20.38$ mag is fainter than the $RF$ value from the IGLS3 \citep{smart13}
by about 1.7 mag (Table~\ref{table2}), suggesting that this object can be highly variable.
Two featureless spectra in SDSS database support its classification as a BL Lac object.

\subsubsection{MST 1020$+$0510}\label{src15} 

There are several FIRST radio sources in the searching region of this cluster, but only 
three of them (whose data are reported in Table~\ref{table2}) have rather safe optical 
counterparts whereas none of them corresponds to known X-ray sources.
The source SDSS J101948.24+051328.8 (note $d$ in Table~\ref{table2}) has the flux density
of 9.5 mJy in FIRST and a featureless SDSS spectrum but a rather high colour index $u - r = 1.7$.
Its WISE colours are uncertain because of a possible confusion with a close starlike object,
but lie quite well within the BL Lac region. 
It appears in the QSO candidate lists by \citet{dabrusco09} and \citet{brescia15}.
The possible optical counterpart SDSS J102011.76+050636.1 (note $e$ in Table~\ref{table2}) of 
the this FIRST radio source (the offset in the position is $\approx$3\arcsec) is a starlike 
object with a large $u - r = 2.9$, without any indication of a blazar appearence.
The source SDSS J102015.12+050910.5 (note $f$ in Table~\ref{table2}) has a radio flux 
density of 10 mJy and corresponds to a rather faint and red galaxy with a large Ca H\&K 
break and a well apparent broad H$\alpha$ line. 
WISE photometry of the latter two sources gives only upper limits in the $W3$ band.
The association of the $\gamma$-ray emission to one of these three objects remains 
quite unsafe.

\subsubsection{MST 1405$-$1853}\label{src16} 

There is only a NVSS radio source inside the searching radius of 6\arcmin.
It has a point-like optical counterpart having all the three WISE colours 
($W1 - W2 = 0.86\pm0.05,~ W2 - W3 = 2.19\pm0.16,~ W3 - W4 = 2.37\pm0.44$) compatible with 
a BL Lac classification according to the WRIBRaLS criteria \citep{dabrusco14}.
It appears, therefore, to be a quite good candidate for an HBL object.

\subsubsection{MST 1410$+$1438}\label{src17} 

\begin{figure*}[ht!]
\centering 
\includegraphics[height=5.2cm,angle=0]{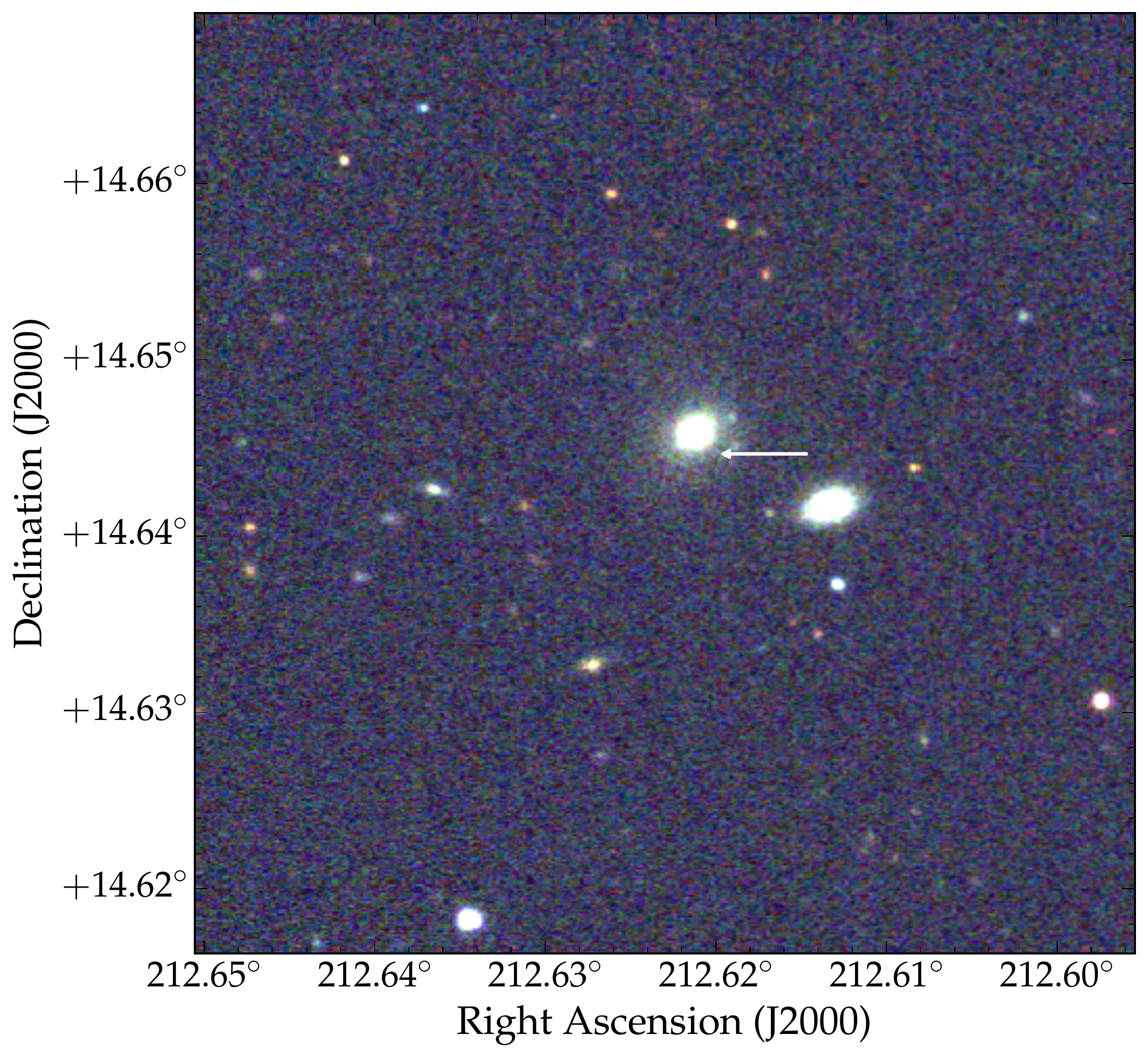}
\includegraphics[height=5.2cm,angle=0]{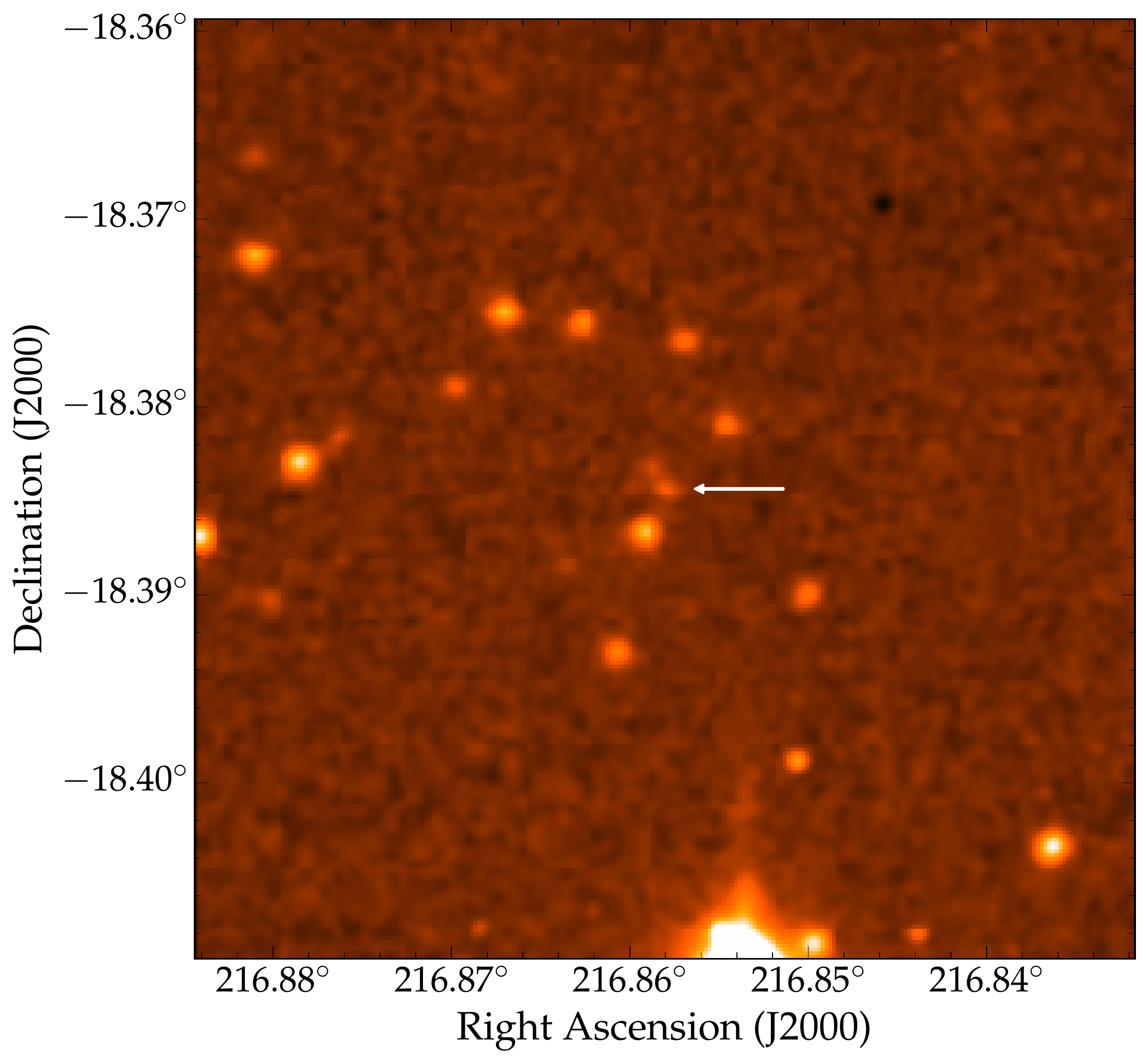}
\includegraphics[height=5.2cm,angle=0]{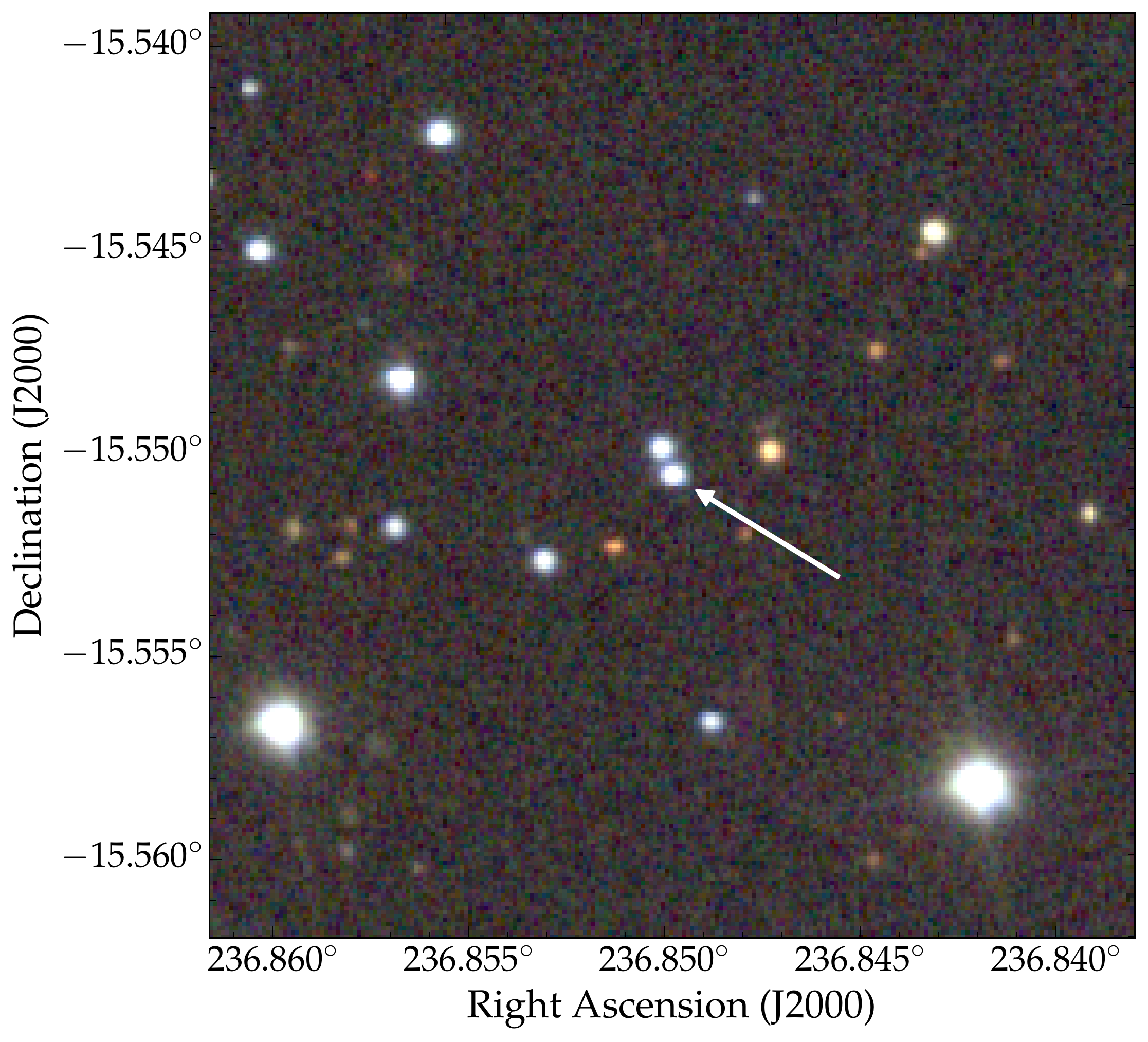}
\caption{\emph{Left panel}: SDSS image ($irg$ composite) of the radio galaxy 
possible counterpart to MST 1410$+$1438 (see Section~\ref{src17}), indicated by the horizontal line.
The image size is 200\arcsec.
\emph{Central panel}: Super Cosmos $J$-band field of the optical counterpart (indicated by the 
horizontal line) of the radio source close to the centroid of MST 1427$-$1823 (Section~\ref{src18}).
The image size is 180\arcsec.
\emph{Right panel}: SDSS image ($irg$ composite) of the close pair one of which could be the counterpart to MST 
1547$-$1531 (Section~\ref{src24}). The image size is 100\arcsec.
The arrow  mark the favorite blazar candidate.
For all panels  North is in the upper direction and East to the left.
}
\label{f2}
\end{figure*}

Very close to the cluster centroid there is a compact radio source with the highest flux
density in our sample and a bordeline [1.4--4.85] GHz spectral index close to $-0.5$, 
but flatter at lower frequencies.
The optical counterpart is one of a couple of galaxies at the same $z = 0.144$ and at an
angular distance of $\approx$30\arcsec\ (see Figure~\ref{f2}, left panel).
It has a low Ca H\&K break ratio ($\approx$0.15), an emission H$\alpha$ line
and the SDSS $r$ is equal to 16.7 mag.
It is reported as a C class $\gamma$-ray 
BL Lac object candidate in WIBRaLS.
According to the 5BZCAT criteria it could be classified as a galaxy-dominated 
blazar.

\subsubsection{MST 1427$-$1823}\label{src18} 

The optical counterpart to the unique radio source in the field is a point-like object
with WISE colours suggesting a blazar candidate (see Figure~\ref{f2}, central panel).
Optical spectroscopy is necessary to confirm this classification.

\subsubsection{MST 1431$-$3122}\label{src19} 

It is the cluster with the highest photon number in the sample.
The PKS and CRATES radio source here indicated as possible counterpart was proposed 
as B class BL Lac and $\gamma$-ray source candidate in WIBRaLS.
It is also detected in the RASS.
The 6dF spectrum is apparently featureless with a possible emission line at the blue
end that, if due to C\textsc{iv}, would imply a redshift $z = 1.632$.
At an angular distance of 1\farcm6 there is also a SUMSS21 source with a flux density 
of 25 mJy, a very faint optical counterpart and a corresponding WISE source.

\subsubsection{MST 1439$-$2524}\label{src20} 

The optical counterpart to the NVSS source is at a few arcseconds from a brigther star
and the only available data are photometry from general surveys.
The WISE colours are close to the blazar area and particularly $W2 - W3$ is the lowest
in all the sample not far from the region of galaxies (orange point in Figure~\ref{f1}).
These poor data does not allow to safely establish its nature and whether it is or not 
the conterpart to the $\gamma$-ray cluster and the possibility of a chance association 
cannot be excluded.

\subsubsection{MST 1503$+$1651}\label{src21} 

Six radio sources are within the searching circle of this cluster, and only the selected 
one has an optical conterpart.
WISE colours ($W1 - W2 = 0.79\pm0.05,~ W2 - W3 = 2.45\pm0.21$) locate this object near 
the centre of the blazar region.
It is in the list of QSO candidatates in WIBRaLS, while the SDSS 
spectrum with a featureless blue continuum strongly supports the HBL classification.

\begin{figure*}[ht!]
\centering 
\includegraphics[height=5.45cm]{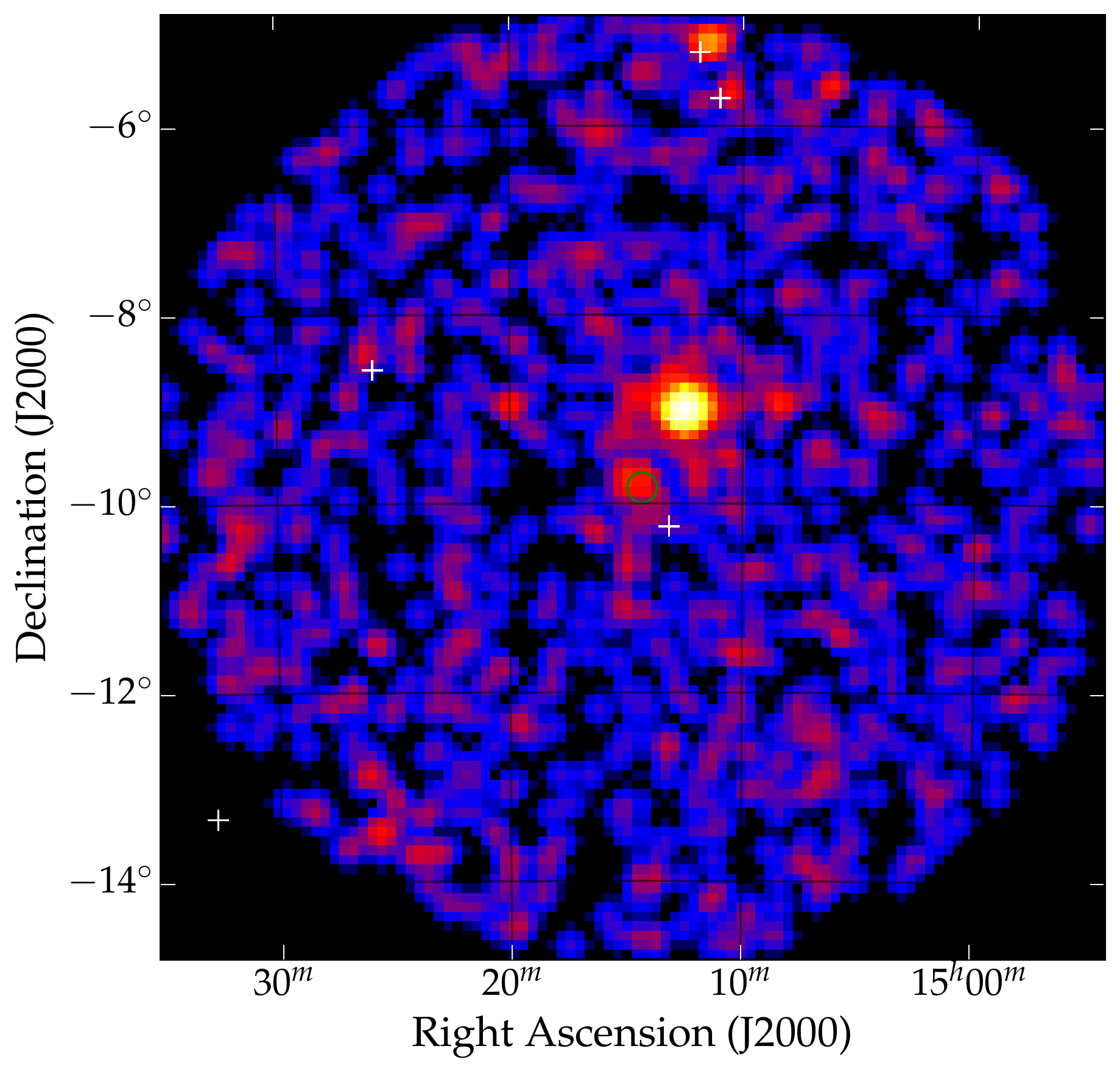}
\includegraphics[height=5.45cm]{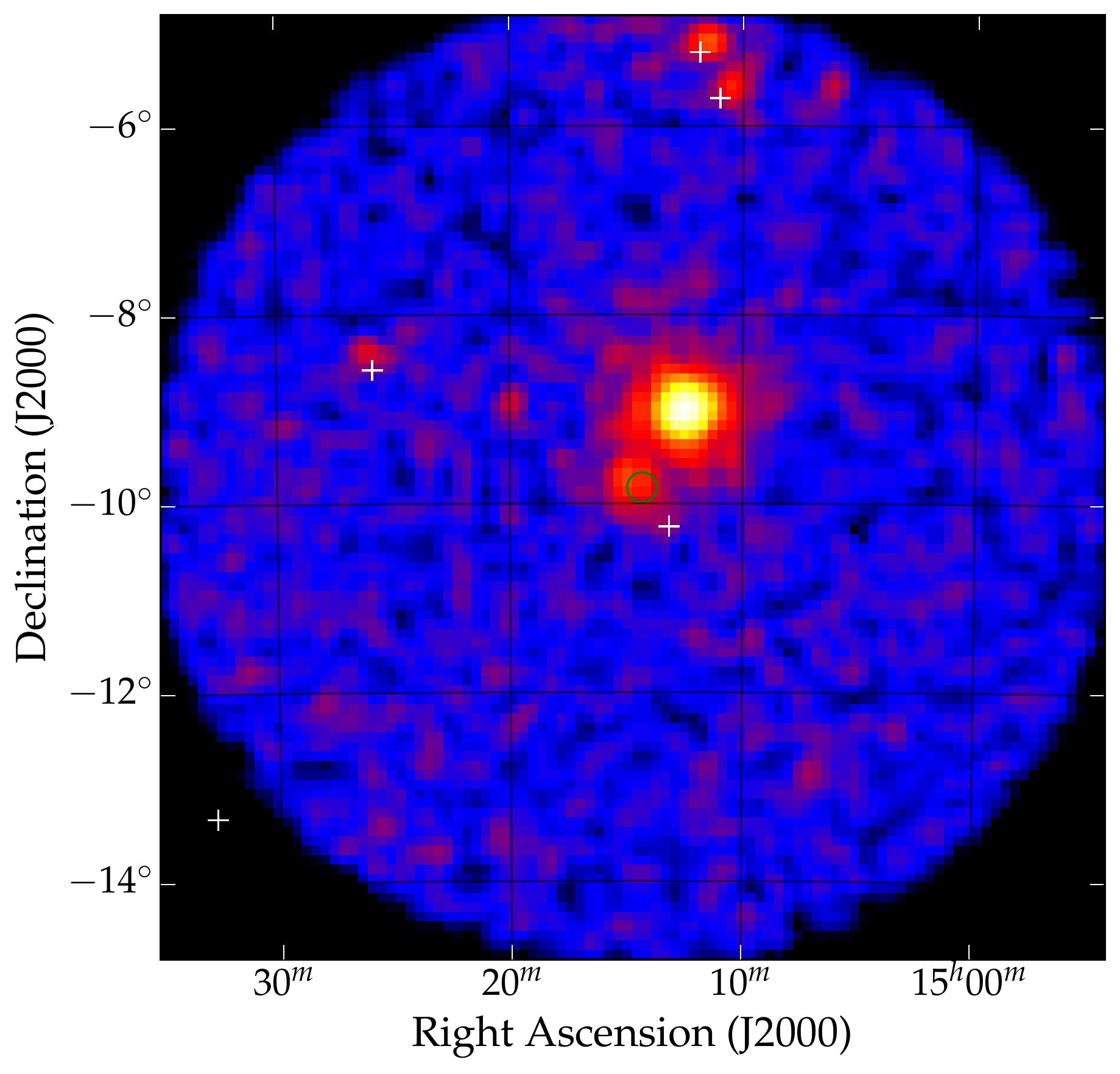}
\includegraphics[height=5.45cm]{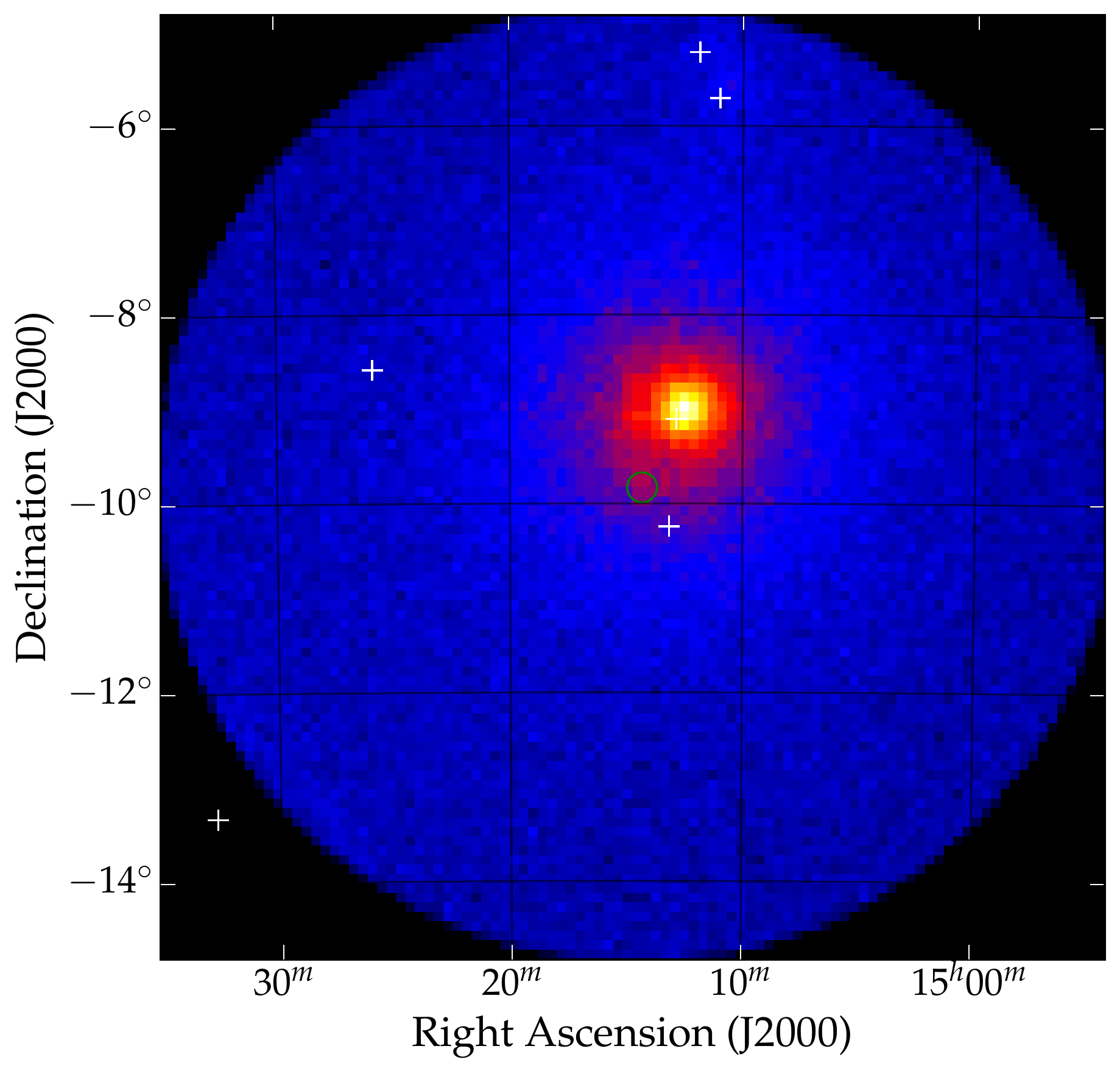}
\caption{False colour count maps in equatorial coordinates at energies higher than
10 GeV (left panel), 3 GeV (central panel), and 100 MeV (right panel), of the sky region 
centred at the cluster MST~1514$-$0949 (green circle). 
The radius of the considered region is 5\degr.
White crosses indicate the positions of 3FGL sources.
See Section~\ref{src22} for details.
}
\label{f3}
\end{figure*}

\subsubsection{MST 1514$-$0949}\label{src22} 

The are some peculiarities on this cluster: it has the lowest $g$ values in the sample 
but it is curiously associated with the highest $\sqrt{TS}$.
An analysis with $\Lambda_\mathrm{cut} = 0.6\,\Lambda_{m}$ found again a significant cluster of 8 
photons with $g = 3.508$ at the same coordinates. 
It should be considered that this cluster is close to the bright source 3FGL J1512.8$-$0906 
($\Delta \theta$ = 0\fdg86), associated with the FSRQ PKS 1510-08 (Abdo et al. 2010c), 
well found in our analysis with $n$ = 278 and $M$ = 1997, and even closer to 3FGL J1513.1-1014
($\Delta \theta$ = 0\fdg56). 
False colour count maps at different energies are shown in Figure~\ref{f3}: a point-like object
can be easily seen above 3 GeV (central panel), and it appears clearly separated from 
the much brighter near source and with a brightness comparable to that of the two 3FGL
sources close to the North boundary of the image.

The possible counterpart is a rather strong radio source with a spectrum from archive data 
that is steep up to about 1 GHz and turns to flat at higher frequencies; it is in the 
CRATES catalogue and in the microwave range it is bright enough to be reported in the 
Planck catalogue with flux densities at 143 and 217 GHz of $\sim$200 mJy. 
In the POSS images it is at about 8\arcsec\ from a much brighter starlike object ($R \approx 16$) 
while that corresponding to the radio source is brighter at WISE wavelengths.
Photometric data are available in all the four WISE bands and the colours place within 
the upper part of the blazar locus where are both BL objects and flat spectrum radio 
quasars.
Its blazar nature should be considered safe but an optical spectrum will be useful for 
establishing the right type.

\subsubsection{MST 1546$-$1002}\label{src23} 

This flat spectrum radio source is reported in the CRATES and LORCAT catalogues.
Its NVSS position corresponds to the fainter object of a pair with a separation of a 
few arseconds.
These objects appears unresolved in the WISE images and a contamination in the photometric 
data cannot be excluded.
In any case their colours are well placed within the blazar region (blue circle in 
Figure~\ref{f1}).
More data will be useful for understanding their nature and to confirm the association
with the $\gamma$-ray cluster.

\subsubsection{MST 1547$-$1531}\label{src24} 

A close pair of objects with similar brightness appear in the SDSS image at the NVSS 
position of this weak radio source (Figure~\ref{f2}, right panel), one of which (SDSS J154725.68-153237.0) 
has $u - r = 0.93$, while the other has a redder color index $u - r = 1.74$.
In the WISE images these two close objects are unresolved and therefore it is not possible 
to determine their individual colours, but it is interesting that their corresponding point 
is close to the center of the BL Lac region (open blue square in Figure~\ref{f1}).
The bluer object appears as a QSO candidate in \citet{brescia15} and should be considered
as the primary target for a blazar search.

\subsubsection{MST 1605$-$1142}\label{src25} 

This source was already reported in the literature as a possible counterpart of a 3EG 
$\gamma$-ray source \citep{tornikoski02,sowardsemmerd04} and was included 
in the CGRaBs list \citep{healey08} but it is not present in all the Fermi-LAT catalogues.
It is also in the LORCAT catalogue \citep{massaro14} and
in the WIBRaLS 
samples.
No optical spectrum is published. 
Our finding of a cluster compatible with its position supports the blazar nature of this
source.  

\subsubsection{MST 1640$+$0640}\label{src26} 

Very few literature data are available on this radio source.
It is reported in the Atlas of Radio/X-ray associations \citep[ARXA,][]{flesch10}
with a high probability of being a QSO.
In the two WISE colours plot it lies inside the blazar region.

\subsubsection{MST 1643$+$3317}\label{s27} 

Five radio sources are in the search radius of this cluster.
Two of them, both having a radio flux density of $\sim$3 mJy, are ``empty field'', 
i.e. without optical counterpart. 
The likely optical counterparts of two other sources, with radio flux densities of 
$\sim$10 and 20 mJy, are two similar faint red galaxies, with SDSS photometry of  
$r = 20.5$ mag (redshift $z = 0.608$), and $r = 20.6$ mag (redshift $z = 0.608$).
More interesting is the counterpart of the brightest radio source (possibly with
two/three components in FIRST) and associated with 1RXS J164339.1+331644, already
considered a candidate QSO \citep{brinkmann97,brescia15}
In the WISE database only the $W1 - W2 = 0.44$ colour is available and its nature 
is not well known because of the lack of an optical spectrum.
On the basis of these data no object exhibits clear blazar properties, and the actual 
counterpart of the $\gamma$-ray cluster remain uncertain.

\begin{figure*}[ht!]
\centering 
\includegraphics[height=5.25cm]{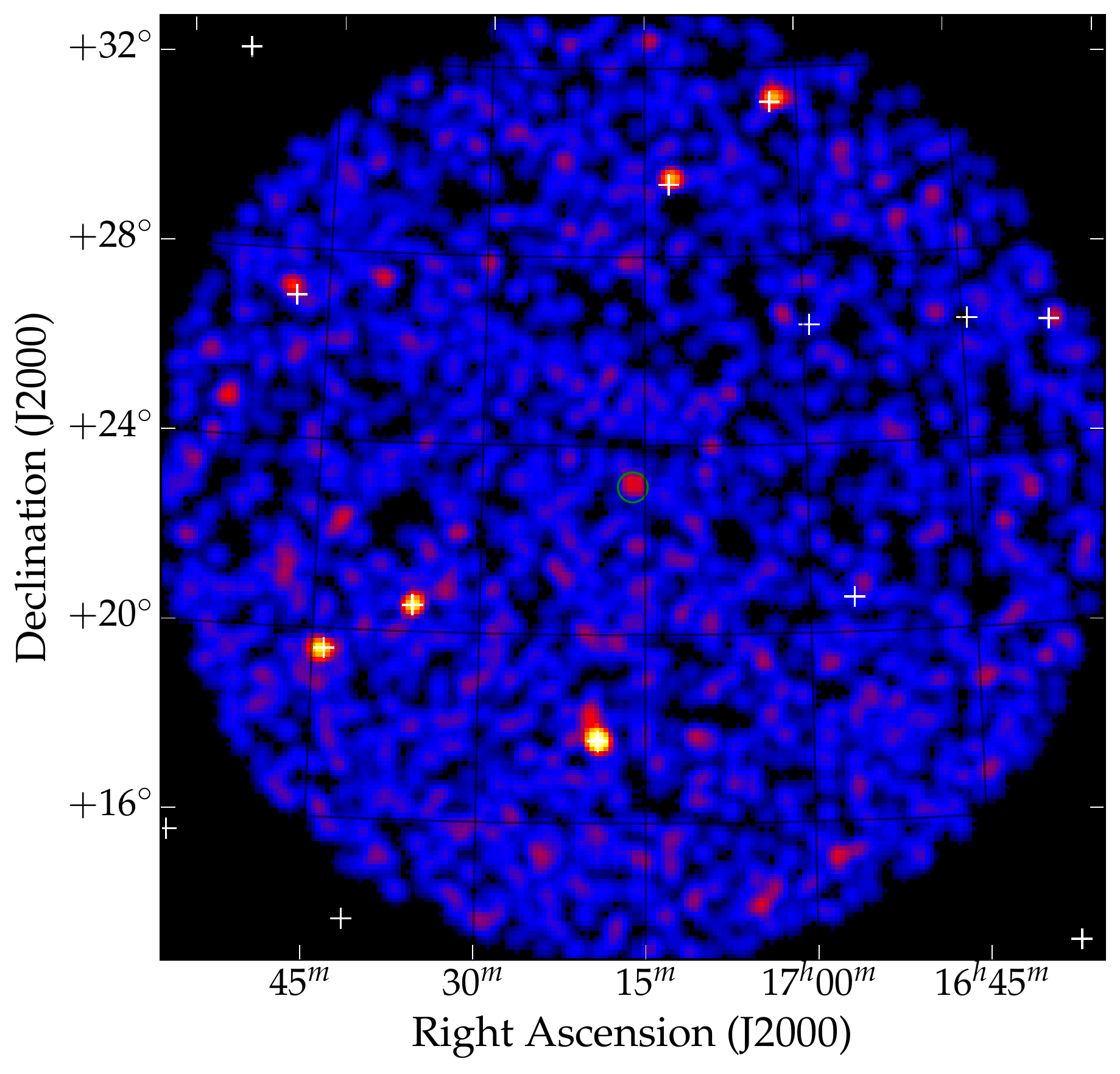}
\includegraphics[height=5.25cm]{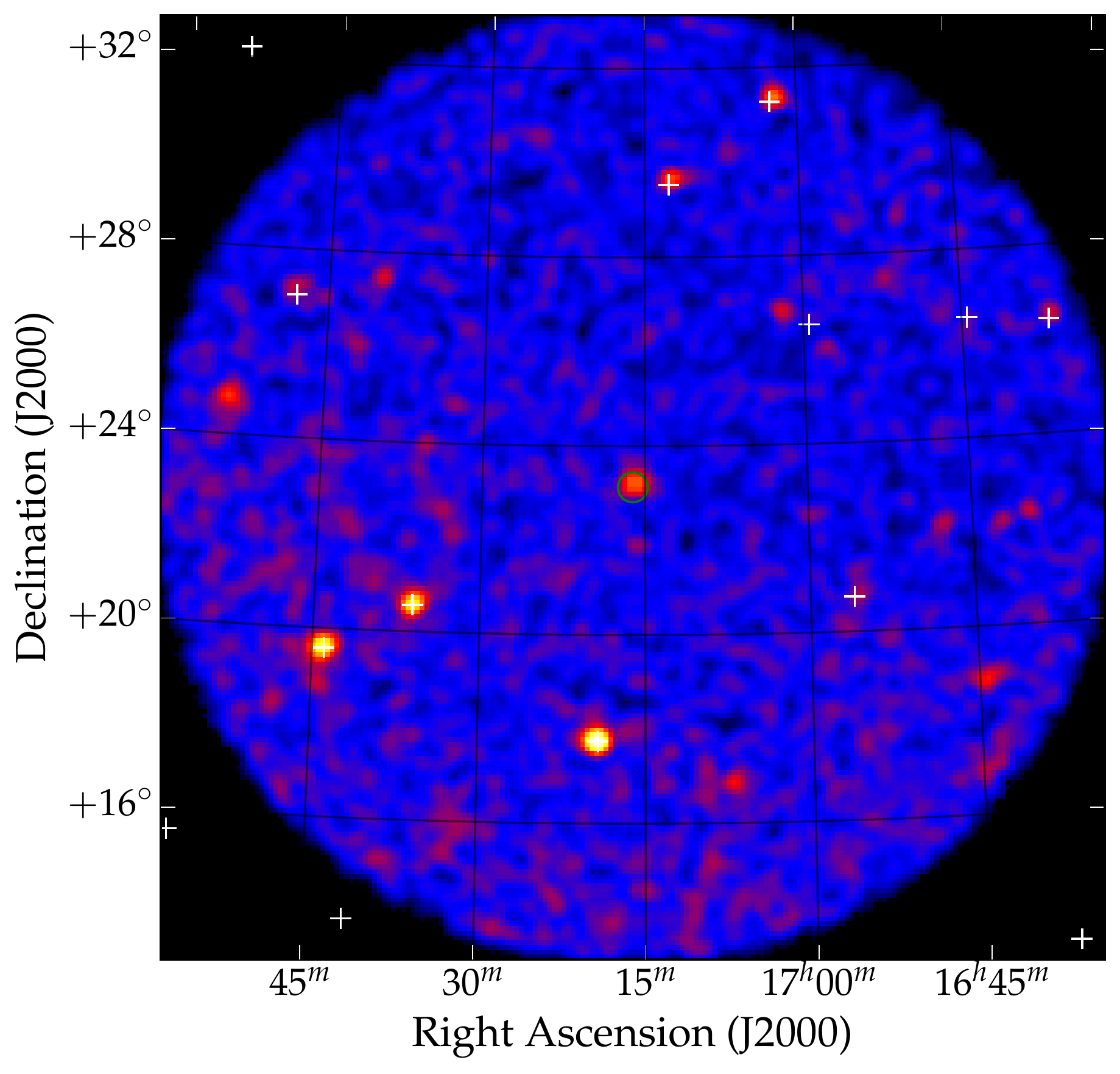}
\includegraphics[height=5.25cm]{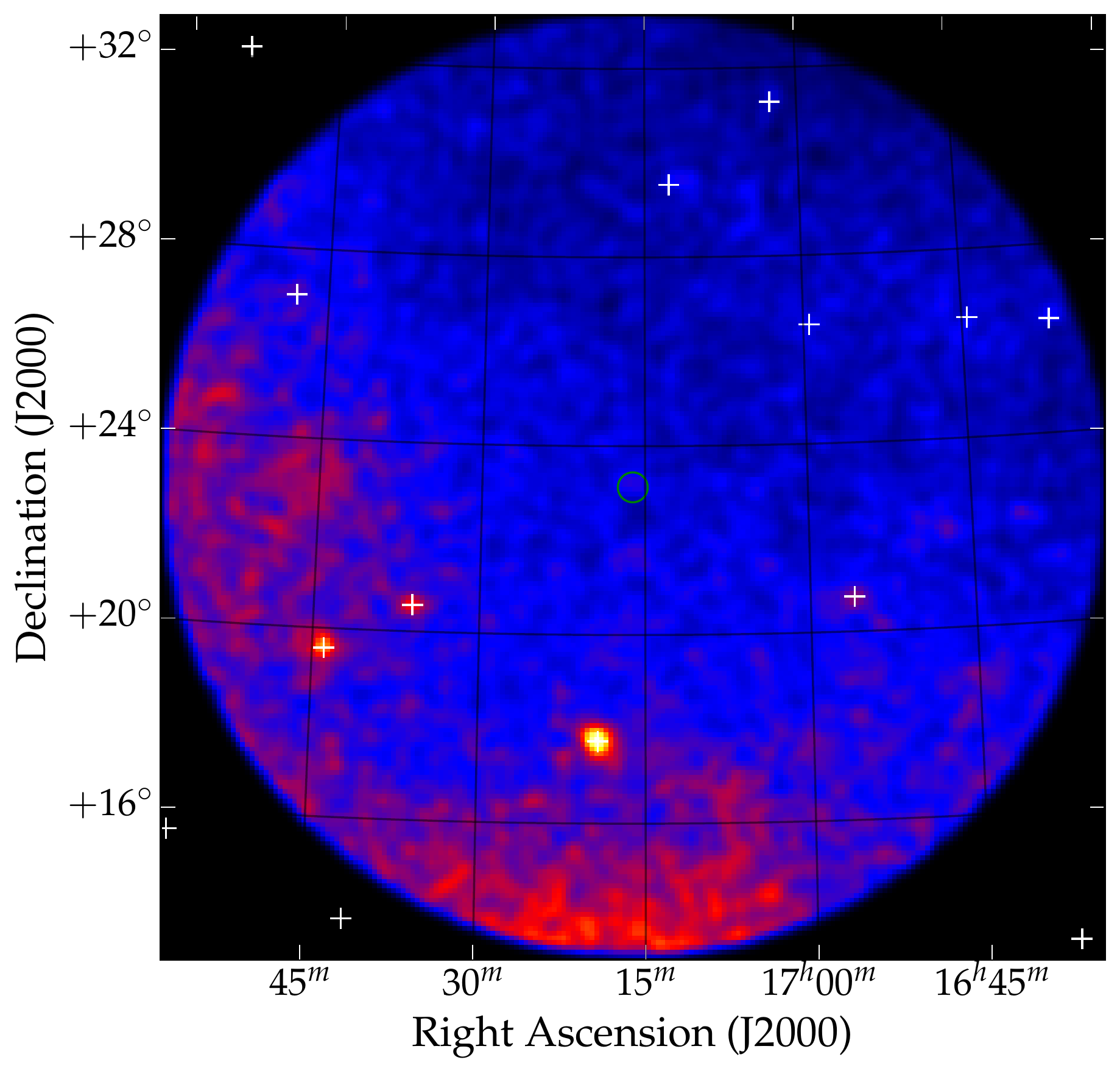}
\caption{False colour count maps in equatorial coordinates at energies higher than
10 GeV (left panel), 3 GeV (center panel) and 100 MeV (right panel) of the sky region 
centred at the cluster MST J1716$+$2307 (green circle). 
The radius of the considered region is 10\degr.
White crosses indicate the positions of 3FGL sources.
See Section~\ref{src28} for details.
}
\label{f4}
\end{figure*}

\subsubsection{MST 1716$+$2307}\label{src28} 

The ML analysis confirmed the existence of a $\gamma$-ray source with the high $\sqrt{TS}=9.7$.
False colour count maps at different energies are shown in Figure~\ref{f4}: a point-like object
can be easily seen above 3 GeV (central and right panel), while it can be confused with the background 
at energies higher than 100 MeV (left panel).
It appears therefore safely established without any possibility of confusion with other sources.
There are two weak radio sources within the matching radius and one of them, at the angular
distance of 2\farcm8, has a very red and faint ($r > 22$) SDSS counterpart, while that of the 
other radio source is a blue object (SDSS J171603.22+230822.8) with a star-like appearance 
and $u - r = 0.87$ similar to other $\gamma$-ray loud HBL objects \citep{massaro12}.
It was associated with an X-ray source in the ARXA \citep{flesch10} and reported as QSO candidate
by \citet{brescia15}.
No optical spectroscopy is available.

\subsubsection{MST 2037$-$3836}\label{src29} 

This cluster is located at a longitude close to that of the Galactic Centre in a region
with a relatively high background.
ML analysis gave for this cluster a significance well below the acceptance threshold 
($\sqrt{TS} = 2.3$).
MST searches at 8 and 5 GeV extracted a cluster with $n = 5$ but with magnitude values lower 
than 20, while at 3 and 20 GeV no cluster was found.
The cluster detection is therefore unstable and depends on the considered region.
Considering the higher background with respect to other sky regions, the probability
that this marginal cluster could be originated by a density fluctuation (the presence
of only one photon is critical for its selection) is not negligible.
In addition, WISE photometric data of the proposed counterpart are uncertain because 
it is close to a much brighter star and is poorly resolved in the IR images.

\subsubsection{MST 2245$-$1734}\label{src30} 

There are three NVSS sources within the searching radius, two of them having a radio flux
density of a few mJy without optical counterparts, while the brightest has a faint possible
counterpart reported as candidate QSO by \citet{brescia15}.
Its WISE colours are rightly located inside the blazar region. 
It appers therefore as a good candidate for a HBL object.

\section{Light curves and variability}

\begin{figure*}[htb]
\centering 
\includegraphics[width=0.47\textwidth]{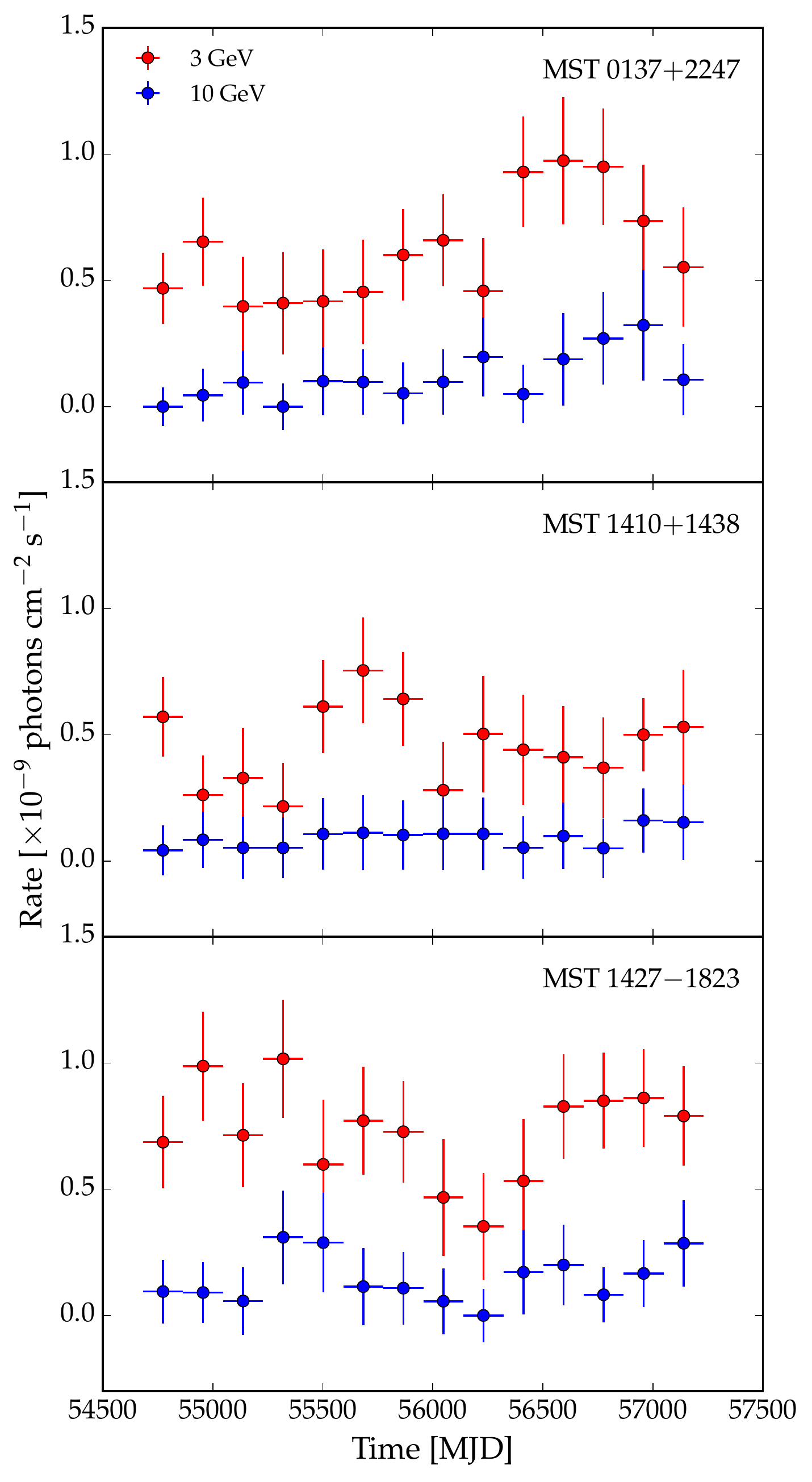}
\includegraphics[width=0.47\textwidth]{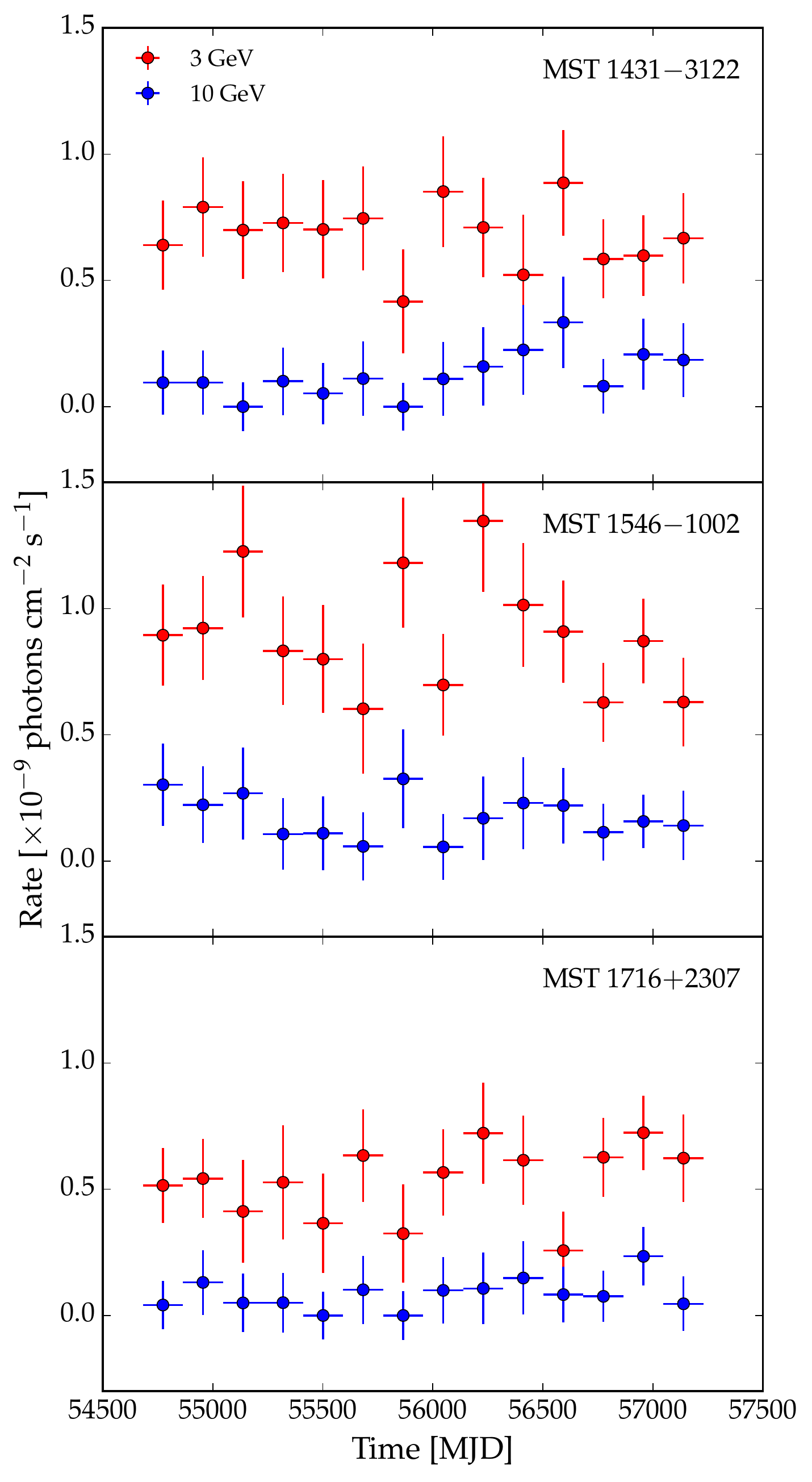}
\caption{
Aperture photometry light curves, in the 3--300 and 10--300 GeV bands, 
extracted with a 1\degr\ radius around the clusters MST~0137$+$2247, 
MST~1410$+$1438, MST~1427$-$1823, MST~1431$-$3122, MST~1546$-$1002 and 
MST~1716$+$2307. See text for details.
}
\label{f5}
\end{figure*}

\begin{figure}[hbt]
\centering 
\includegraphics[width=0.48\textwidth]{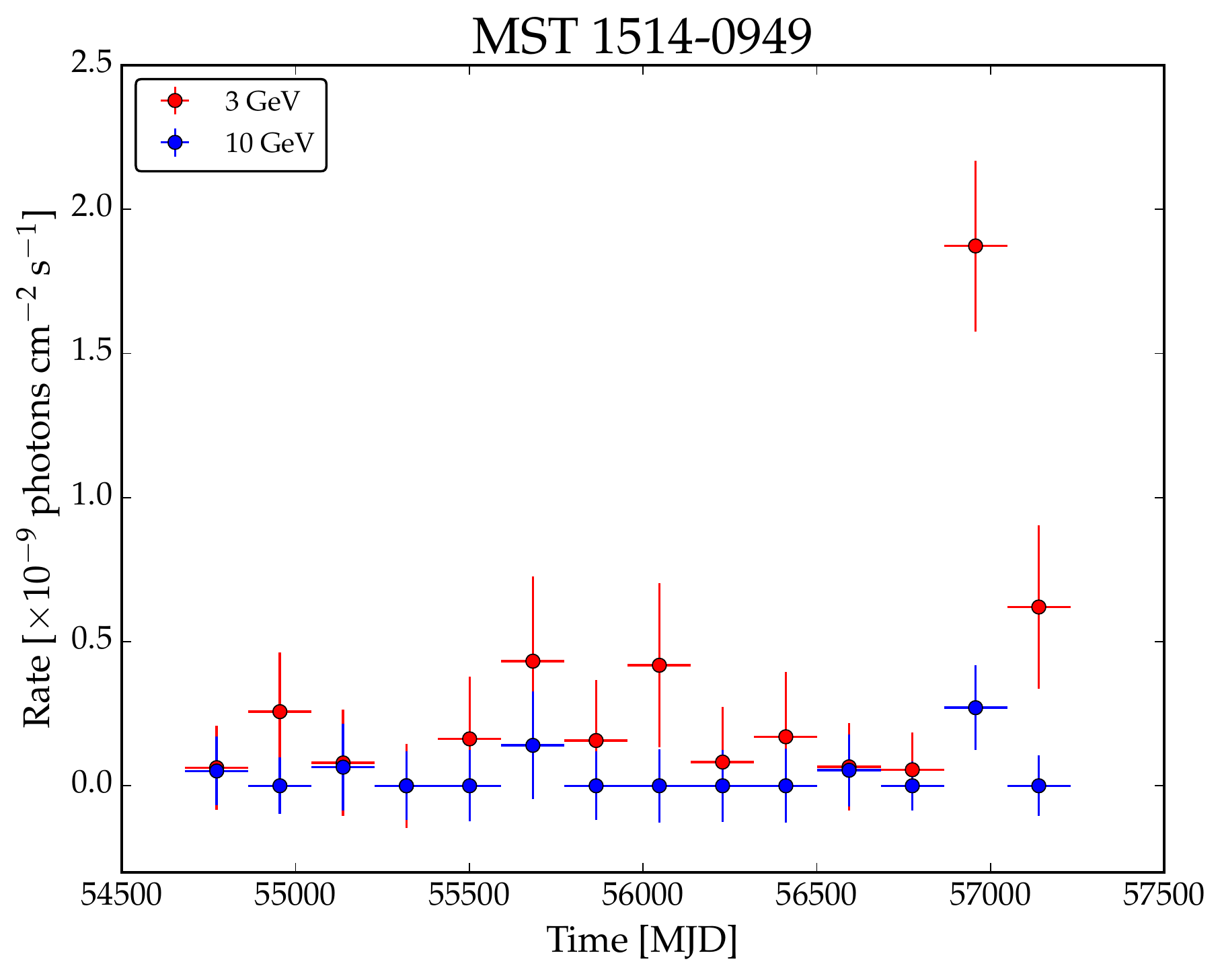}
\caption{
Aperture photometry light curves, in the 3--300 and 10--300 GeV bands, 
extracted with a 0\fdg25 radius around the cluster MST~1514$-$0949. See text for details.}
\label{f6}
\end{figure}

Blazars are variable objects and therefore, for the seven clusters 
for which we derived a flux higher than 10$^{-10}$ photons cm$^{-2}$ s$^{-1}$ in the energy 
band 3--300 GeV, we extracted light curves by means of standard aperture 
photometry\footnote{\url{http://fermi.gsfc.nasa.gov/ssc/data/analysis/scitools/aperture_photometry.html}}, 
with an extraction radius of 1\degr.  
For the source MST 1514$-$0949 (Section~\ref{src22}), due to the possible contamination 
from the nearby sources PKS 1510-08 and 3FGL J1513.1-1014, the extraction radius was 
reduced to 0\fdg25.
Light curves with a 6-months binning and in the 3--300 and 10--300 GeV bands for 
these clusters are shown in Figures~\ref{f5} 
and \ref{f6}.

For some sources there is a hint of variability, especially in the 3--10 GeV band, 
on months to years timescales. 
For example, MST~0137$+$2247 showed a flux increase around MJD 56600--57000, and for
MST 1514$-$0949 there is evidence of a single, bright flare around MJD 57000.
The possibility of contamination from the nearby PKS 1510-08 was excluded by 
extracting and comparing a light curve for the latter source with a narrow 0\fdg5 radius.

\section{Summary and discussion}

We analysed the first 7 years of \emph{Fermi}-LAT sky at energies higher than 10 GeV 
by means of the MST algorithm, allowing a robust detection of photon 
clusters having typical sizes comparable with the instrumental point spread function.
In the selection procedure we adopted rather severe threshold values to reduce 
the possibility of spurious detections due to local background fluctuations.

In the present paper we report 30 new clusters, 26 of them were fully confirmed by 
the ML analysis that gave $\sqrt{TS}$ higher than 5; for two clusters we obtained 
values higher than 4.5, while the remaining two have too low $TS$ to be considered
as confirmed.
These clusters were also associated with blazar candidates selected on the basis of
radio and optical detections.
For few of them 6dF and SDSS spectra are available, 
while some sources appear to be low redshift radio galaxies.
The nature of the other sources must be confirmed by new spectroscopic observations.

The blazar nature of our candidates is also supported by their WISE two-colour
plot, in which all selected sources are in the blazar region, with a clear dominance
for that of BL Lacs.
It appears quite unlikely that radio sources, with likely counterparts in the high 
energy $\gamma$-rays, exhibit mid-IR colours typical of BL Lacs and do not belong
to this class of AGNs.
These researches extend the knowledge on the BL Lac population in two directions:
one is towards low brightness sources and the other is concerning the existence
of a subclass of BL Lacs too faint in radio band to be detected in the  available
surveys.
We think that a systematic search for some possible counterparts of $\gamma$-ray
clusters with $M$ values lower than the threshold considered here and based on
a multifrequency approach, particularly in the mid IR band, could be useful to
enrich these studies with new observational results. 

Considering the results of Papers I, II, and III, together with those of the present
work, to now we have found evidence for 90 blazar or candidate selected for their
emission above 10 GeV detected by means of MST cluster search.
We remark, however, that these discoveries are due not only to our clustering method 
but mainly to the improvement of instrumental response functions used for producing 
the Pass 8 sky and to about double exposure duration with respect to that considered 
at the epoch of 3FGL catalogue.

\begin{acknowledgements}

We acknowledge use of archival Fermi data. We made large use of the online version of the Roma-BZCAT 
and of the scientific tools developed at the ASI Science Data Center (ASDC),
of the final release of 6dFGS archive,
of the Sloan Digital Sky Survey (SDSS) archive, of the NED database and other astronomical 
catalogues distributed in digital form (Vizier and Simbad) at Centre de Dates astronomiques de 
Strasbourg (CDS) at the Louis Pasteur University.
\end{acknowledgements}

\bibliographystyle{spr-mp-nameyear-cnd}
\bibliography{bibliography} 

\end{document}